\documentclass[reprint,amsmath,amssymb,aps,prl]{revtex4-2}
\usepackage{xcolor}
\usepackage{graphicx}

\usepackage{chngcntr}

\begin{document}

\title{Optimal rigid brush for fluid capture}
\author{Basile Radisson}
\author{Hadrien Bense}
\author{Emmanuel Si\'efert}
\author{Lucie Domino}
\author{Hoa-Ai B\'eatrice Hua}
\author{Fabian Brau}\email{fabian.brau@ulb.be}
\affiliation{Universit\'e libre de Bruxelles (ULB), Nonlinear Physical Chemistry Unit, CP231, 1050 Bruxelles, Belgium}

\date{\today}

\begin{abstract}
Parallel assemblies of slender structures forming brushes are common in our daily life from sweepers to pastry brushes and paintbrushes. This type of porous objects can easily trap liquid in their interstices when removed from a liquid bath. This property is exploited to transport liquids in many applications ranging from painting, dip-coating, brush-coating to the capture of nectar by bees, bats and honeyeaters. Rationalizing the viscous entrainment flow beyond simple scaling laws is complex due to its multiscale structure and the multidirectional flow. Here, we provide an analytical model, together with precision experiments with ideal rigid brushes, to fully characterize the flow through this anisotropic porous medium as it is withdrawn from a liquid bath. We show that the amount of liquid entrained by a brush varies non-monotonically during the withdrawal at low speed, is highly sensitive to the different parameters at play and is very well described by the model without any fitting parameter. Finally, an optimal brush geometry maximizing the amount of liquid captured at a given retraction speed is derived from the model and experimentally validated. These optimal designs open routes towards efficient liquid manipulating devices.
\end{abstract}

\maketitle

Many strategies exist to transport fluids at scales of the order of the capillary length. Passive transport, driven by surface tension, has proven to be an effective mean to promote directional liquid transport provided the surface is adequately structured~\cite{Chen2016,Feng2021} with practical applications in microfluidics~\cite{Seemann2012,Stone2004} or water harvesting~\cite{Zheng2010,Park2016}. Capillary rise within the interstices of an assembly of structures, both when the system is rigid~\cite{Princen1969,Charpentier2020} or deformable by capillary forces~\cite{bico2004elastocapillary,Kim2006,py20073d,duprat2011dynamics}, is another method to control the transport of liquid. According to the geometry of the system and the viscosity of the liquid, the capillary rise can be relatively small or slow. Active transport is then a way to capture a larger amount of fluid more quickly and can involve a range of mechanisms, from pressure difference to ``suck'' a liquid~\cite{Wei2023}, to viscous entrainment. An archetypal example of the latter is dip-coating where an immersed object is pulled out of a liquid bath. The importance of this process in many industrial applications is illustrated by the vast literature which, since the seminal work of Landau, Levich and Derjaguin~\cite{Landau1942,Derjaguin1943}, explored many instances of dip coating~\cite{Quere1999,Weinstein2004,Tang2017,Bertin2022}. In particular, textured flat surfaces and rods were shown to enhance the collect of liquid by viscous entrainment in one-dimensional settings where the flow is mainly unidirectional~\cite{seiwert2011coating,nasto2018viscous,lechantre2019collection,cheng2023viscous}. In the case of flexible hair bundles, the retraction speed increases the capillary attraction force between neighboring hairs~\cite{bense2023measurement} and affects the morphology of the bundle itself due to the interplay between capillary and viscous forces~\cite{ha2020hydrodynamic,Moon2024} but the impact of flexibility on the liquid transport remains to be elucidated.

\begin{figure*}[!t]
\centering
\includegraphics[width=\textwidth]{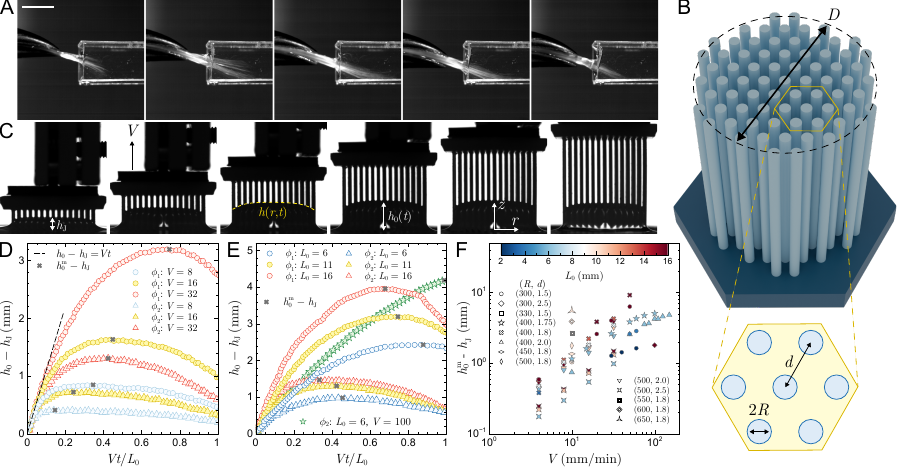}
\caption{(A) Snapshots of a honeyeater (\textit{Acanthagenys rufogularis}) feeding on sugar solution (scale bar: 5 mm; $\Delta t = 10$ ms). Credit: A. E. Hewes and A. Rico-Guevara. (B) Schematic of a brush composed of an equilateral triangular array of pillars showing its lateral size $D$, the distance $d$ between the centers of neighboring pillars and their radius $R$. (C) Snapshots of an experiment where a brush ($R=300$ $\mu$m, $d=1.5$ mm) initially immersed at a depth $L_0 = 16$ mm is withdrawn from a bath of silicon oil ($\mu = 0.97$ Pa s) at a retraction speed $V = 30$ mm/min ($\Delta t = 6.4$ s). The height $h$ of the interface is initially equal to the Jurin's height $h_J$ and varies during the retraction. This evolution is monitored by measuring the temporal evolution of interface height at the center of the brush, $h_0(t) = h(0,t)$. (D) Evolution of $h_0-h_J$ as a function of time, rescaled by the time needed to displace the brush by a distance $L_0=11$ mm, for two brushes ($R=500$ $\mu$m, $d=2.0$ mm [$\phi_1 = 0.773$] and $R=500$ $\mu$m, $d=2.5$ mm [$\phi_2 = 0.855$]) withdrawn at various speed $V$ as indicated in mm/min. The maximum height $h_0^m$ reached by the interface during the retraction process is indicated with cross symbols. The dashed curve indicates a motion at the speed of the brush. (E) Same as panel (D) for various immersion depth $L_0$ as indicated in mm and $V=32$ mm/min except for the data represented by a star symbol where $V = 100$ mm/min. (F) Evolution of $h_0^m-h_J$ as a function of retraction speed for various brushes and immersion depth ($\mu = 0.97$ Pa s).}
\label{fig01}
\end{figure*}

Dipping brush-like structures is also a strategy adopted by some nectarivores to feed on nectar~\cite{Inouye2013,kim2011optimal}. Indeed, collecting a viscous fluid at small scales prevents the use of methods employed by other animals~\cite{kim2012natural}, like using gravity (humans) or fluid inertia to overcome gravity (lapping for cats~\cite{reis2010cats}, ladling for dogs~\cite{crompton2011dogs,gart2015dogs}). To deal with capillary and viscous forces dominating at small scale, many nectarivores have developed highly specialized mouthparts adapted to their feeding method~\cite{Krenn2005,Krenn2019}: hollow tubular proboscis/tongue for suction (butterfly~\cite{Krenn2010}, sunbird~\cite{Paton1989,Cuban2024}) or tongue decorated by numerous outgrowths resembling a brush for dipping (bees~\cite{lechantre2021essential,Wei2023}, honeyeaters~\cite{Mitchell1990,Hewes2023}, bats~\cite{Harper2013}). In the latter case, the tongue is dipped cyclically into the nectar which is collected by viscous entrainment when the tongue is withdrawn from the liquid [Fig~.\ref{fig01}A]. 

Dipping a brush appears thus as a simple and commonly used method for capturing liquids in many contexts, yet the details of the flows occuring in the system, that determine the amount of liquid that can be captured by a brush, remain unclear. Indeed, so far, the rationalization of viscous entrainment in brush-like structures is limited to scaling laws or unidirectional flows. A comprehensive modeling of the flow in these systems thus appears as a necessary step to fully exploit them and to better understand, for example, the physics of dipping among nectarivores.

Here, we use rigid brush structures to study the capture of liquid by viscous entrainment and analyze the nontrivial motion of the air-liquid interface within the brush during its retraction from a liquid bath [Fig.~\ref{fig01}B,C]. Drawing an analogy between brushes and porous media, we derive an analytical model to characterize the 3D flow within a brush during its withdrawal. This model accurately describes the evolution of the interface during an experiment and hence the amount of liquid entrained at any time. Finally, building on our model, we find an optimal brush geometry maximizing the volume of liquid collected at the end of retraction, which is experimentally validated. Our model sheds light on the hydrodynamics of brush-like structures and provides a new tool to design optimal structures for fluid transport.

\textbf{Experiments.} We study the collection of liquid by viscous entrainment using 3D-printed brushes with a mean diameter $D$ and composed of an equilateral triangular array of pillars of radius $R$ separated by a distance $d$ [Fig.~\ref{fig01}B]. The brushes are clamped to a traction device and immersed at a depth $L_0$ in silicon oil of viscosity $\mu$, surface tension $\gamma$ and density $\rho$. Upon immersion, the liquid first rises by capillarity up to the Jurin's height given by~\cite{Princen1969,Charpentier2020}
\begin{equation}
\label{jurin-hex}
h_J = 2\ell_c^2 R^{-1}(1-\phi)\phi^{-1}\, \cos \theta_Y,
\end{equation}
where $\phi = 1-2\pi R^2/(\sqrt{3}d^2)$ is the porosity, $\ell_c = [\gamma/\rho g]^{1/2}$ the capillary length, and $\theta_Y$ the Young contact angle. After the equilibrium state has been reached, the brush is removed at a constant speed $V$ from the bath and the spatio-temporal evolution of the air-liquid interface height inside the brush, $z=h(r,t)$, is recorded from the side with a camera (Fig.~\ref{fig01}C, see Methods for more details). 

Figure~\ref{fig01}D shows some typical temporal evolution of the interface height measured at the center of the brush, $h_0(t)=h(0,t)$, when the retraction speed $V$ is varied while keeping the immersion depth $L_0$ constant (Movies S1 and S2). The interface first moves up at the same speed as the pillars (dashed curve in Fig.~\ref{fig01}D), before slowing down to reach a maximum value $h_0^m$ where it stops and starts moving down. The higher the retraction speed $V$, the more liquid is entrained and the more $h_0^m$ drifts towards the end of the experiments (i.e. $Vt/L_0=1$). Consequently, when $V$ is large enough, $h_0^m$ is reached at the end of the retraction and the temporal evolution of $h_0$ is monotonic as seen in Fig.~\ref{fig01}E (green star symbols). Similarly, increasing $L_0$ while keeping $V$ constant yields larger values of $h_0^m$ [Fig.~\ref{fig01}E] until it saturates at large enough $L_0$ (Movie S3 and SI Appendix).

The fluid capture dynamics by a porous brush results from a competition occurring during the retraction between gravity draining the fluid out of the brush and viscous forces opposing to this drainage. The typical draining time is given by Darcy's law~\cite{Guyon2015} and writes as $t_{\text{D}}=L_0/ V_{\parallel}$ where $V_{\parallel} = k_{\parallel}\rho g/\mu$ is the speed at which a liquid of viscosity $\mu$ flows vertically inside a porous medium of permeability $k_{\parallel}$ due to the gravitational acceleration $g$. This time has to be compared to the duration of the retraction $t_{\text{exp}}=L_0/V$, yielding a first dimensionless parameter $\overline{V}=t_{\text{D}}/t_{\text{exp}} = V/V_{\parallel}$. When $\overline{V} \rightarrow 0$ the fluid has time to flow out of the brush and there is no liquid capture. Conversely, when $\overline{V} \rightarrow \infty$ the fluid that resides inside the brush at $t=0$ does not have time to flow out and moves up with the pillars at speed $V$.

A second dimensionless parameter is identified by noting that, during the retraction, part of the fluid initially inside the brush escapes by flowing longitudinally through the bottom of the brush and transversely through the sides. According to Darcy's law, the flow rate in each direction is given by $Q_i \sim k_i A_i \Delta p_i /(\mu \ell_i)$ where $k_i$ is the permeability of the brush in the considered direction, $A_i$ the cross-sectional area of the flow ($A_{\parallel} \sim D^2$, $A_{\perp} \sim D L_0$) and $\Delta p_i$ the pressure difference that drives the flow over the length $\ell_i$ ($\ell_{\parallel} \sim L_0$, $\ell_{\perp} \sim D$). Therefore, $Q_{\perp}/Q_{\parallel} \sim \bar{\delta}^2 \Delta p_{\perp}/\Delta p_{\parallel}$ where $\bar{\delta}= (k_{\perp}/k_{\parallel})^{1/2} (2L_0/D)$ is the dimensionless parameter that compares the radial to vertical flow rate for a given pressure difference. If $\bar{\delta} \rightarrow 0$, the flow occurs in the longitudinal direction whereas if $\bar{\delta} \rightarrow \infty$, the fluid escapes only through the sides.

However, this argument does not hold above the bath level where the fluid cannot escape the brush transversely because of the presence of the air-liquid interface. The flow is thus essentially longitudinal in this region. The amount of fluid initially above and below the bath level scales as $\phi D^2 h_J$ and $\phi D^2 L_0$, respectively. The ratio between these two quantities, $\bar{h}_J= h_J/L_0$, is another measure of the importance of the transverse flow in the system. When $\bar{h}_J \gg 1$, the liquid initially inside the brush is mostly above the bath level, hence the flow is primarily longitudinal (even for large $\bar{\delta}$).

Figure~\ref{fig01}F gathers all our experiments and shows the evolution of the maximal height $h_0^m$ as a function of the retraction speed for various brushes and immersion depths. Apart from the global increase of $h_0^m$ with the retraction speed, there is no clear trend as shown by the variation of $h_0^m$ with $L_0$ at a given retraction speed. The interplay between the retraction speed, the immersion depth and the porosity of the brush, encoded in the 3 dimensionless parameters identified above, yields intricate results that require a theoretical model to be rationalized.

In the following, we perform a formal analysis of the flow occurring inside a brush withdrawn at constant speed from a bath and obtain a nonlinear partial differential equation for the spatio-temporal evolution of $h(r,t)$ involving the three dimensionless parameters $\overline{V}$, $\bar{\delta}$ and $\bar{h}_J$ identified above. The predictions of this model show quantitative agreement with experimental results obtained with 3D-printed brushes as those reported in Fig.~\ref{fig01}D, E. Finally, we exploit the understanding obtained from this theoretical analysis to identify optimal brushes that maximize the fluid intake.

The formalism developed here is easily adapted to any porous media in which the drainage flow occurs both vertically and horizontally provided the appropriate permeability $k_{\parallel}$ and $k_{\perp}$ are known. This broad applicability is demonstrated by the quantitative agreement between our model and the data obtained with two parallel plates separated by a small gap and withdrawn from a bath (see SI Appendix).

\begin{figure*}[!t]
\centering
\includegraphics[width=\textwidth]{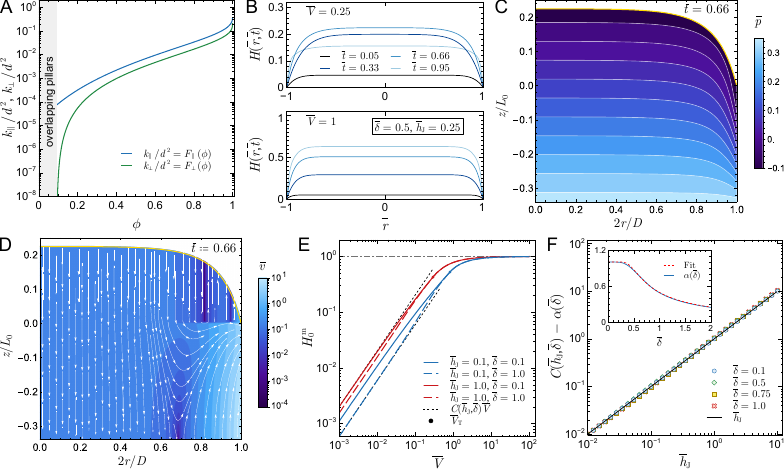}
\caption{(A) Evolution of the longitudinal and transverse permeabilities rescaled by $d^2$ as a function of the porosity $\phi$ for an equilateral triangular array of pillars where $4F_{\parallel} = [1-0.91 \sqrt{x}]^{4.6}[1-0.069(\ln x + K - 35 x)]$ and $12.4F_{\perp} = [1-1.05\sqrt{x}]^{2.5}[1-0.21(\ln x+K)]$ with $x = 1-\phi$ and $K=1.498$. These expressions are fits of numerical data (see SI Appendix) and extend analytical asymptotic expressions~\cite{Drummond1984,Jackson1986}. (B) Spatio-temporal evolution of $H(\bar{r},\bar{t})$, computed with Eq.~(\ref{eq-H-adim}), as a function of time for two retraction speeds, $\overline{V}$, and fixed values of $\bar{\delta}$ and $\bar{h}_J$. Pressure field $\bar{p}=p/\rho g L_0$ (C) and velocity field normalized by $V$ (D) at $\bar{t}=0.66$ for $\overline{V}=0.25$, $\bar{\delta} = 0.5$ and $\bar{h}_J=0.25$ (see top panel (B)). The color map in panel (D) corresponds to the norm of the velocity $\bar{v}=|\vec{v}|/V$. (E) Theoretical evolution of the largest value of $H_0$ in the time interval $0 \le \bar{t} \le 1$, $H_0^m$, as a function of $\overline{V}$ for several values of $\bar{\delta}$ and $\bar{h}_J$. The dots on the curves indicate the value $\overline{V}_{T}$ of the retraction speed beyond which $H_0^m$ is reached at $\bar{t}=1$. At low $\overline{V}$, $H_0^m$ varies linearly with $\overline{V}$, $H_0^m = C(\bar{h}_J,\bar{\delta}) \overline{V}$. (F) Evolution of $C(\bar{h}_J,\bar{\delta})$ as a function of $\bar{h}_J$ for several values of $\bar{\delta}$. The numerical data are fitted by $C = \bar{h}_J + \alpha(\bar{\delta})$. Inset: Evolution of $\alpha$ as a function of $\bar{\delta}$ together with the fit $\alpha^2(x)\approx \tanh[1/(2 x)^2]$.}
\label{fig02}
\end{figure*}

\textbf{Model.} The velocity field in a brush is given by Darcy's law in cylindrical coordinates whose origin is placed in the middle of the brush at the bath level [Fig.~\ref{fig01}C]
{\small
\begin{equation}
\label{vr-vz}
v_{r}(r,z) = -\frac{k_{\perp}(z)}{\mu} \frac{\partial p}{\partial r}, \quad
v_z(r,z) = V - \frac{k_{\parallel}}{\mu} \left(\frac{\partial p}{\partial z} + \rho g\right),
\end{equation}
}
where $V$ is the retraction speed, $k_{\parallel}$ the longitudinal permeability along the $z$-axis and $k_{\perp}(z) = k_{\perp} \theta(-z)$ an effective transverse permeability along the $r$-axis, where $\theta(z)$ is the Heaviside function, allowing an horizontal flow in the immersed part of the brush only. The expressions of $k_{\parallel}$ and $k_{\perp}$ are given in Fig.~\ref{fig02}A (see SI Appendix for more details). Mass conservation, $\nabla \cdot \mathbf{v}=0$, gives an equation for the pressure
\begin{equation}
\label{eq-p}
\frac{\partial^2 p}{\partial z^2} +\frac{k_{\perp}(z)}{k_{\parallel}}\nabla_r^2 p =0,
\end{equation}
where $\nabla_r^2 p = r^{-1}\partial_r(r \partial_r p)$ is the radial part of the Laplacian in cylindrical coordinates. Once the pressure field is known, Eq.~(\ref{vr-vz}) yields the velocity field which, evaluated at air-liquid interface $z=h(r,t)$, gives the spatio-temporal evolution of the interface from the kinematic condition
\begin{equation}
\label{h-eq}
\partial h(r,t)/\partial t = v_z(r,h).
\end{equation}

Introducing the following rescaled quantities
{\small
\begin{equation}
\label{change-var}
\bar{r} = 2r/D, \ (\bar{z},\bar{h},\bar{L},\bar{h}_{J},\bar{t},\bar{p}) = (z,h,L,h_J,V t,p/\rho g)/L_0,
\end{equation}
}
Eq.~(\ref{eq-p}) becomes
\begin{equation}
\label{eq-p-adim}
\frac{\partial^2 \bar{p}}{\partial \bar{z}^2}+ \delta^2 \frac{k_{\perp}(\bar{z})}{k_{\parallel}} \nabla_{\bar{r}}^2 \bar{p} =0, \quad \delta = \frac{2L_0}{D},
\end{equation}
where the system aspect ratio $\delta$ is assumed to be small. Expanding the pressure as a power series in $\delta$
\begin{equation}
\label{p-expan}
\bar{p}(\bar{r},\bar{z}) = \bar{p}_0(\bar{r},\bar{z}) + \delta^2\, \bar{p}_1(\bar{r},\bar{z})+ \ldots,
\end{equation}
Eq.~(\ref{eq-p-adim}) gives at order $\delta^0$ and $\delta^2$
\begin{equation}
\label{eq-p0-p1}
\frac{\partial^2 \bar{p}_0}{\partial \bar{z}^2}=0, \quad \frac{\partial^2 \bar{p}_1}{\partial \bar{z}^2}=- \frac{k_{\perp}(\bar{z})}{k_{\parallel}} \nabla_{\bar{r}}^2 \bar{p}_0.
\end{equation}
To solve Eq.~(\ref{eq-p0-p1}), we assume that the pressure is hydrostatic below the immersed part of the brush, $p(r,-L) = \rho g L$, and equal to the capillary pressure at the air-liquid interface, $p(r,h) = -\rho g h_J$. Here, $L = L_0-Vt$ is the immersed length that decreases linearly in time as the brush is removed vertically from the bath at a constant speed $V$. Using the rescaled variables (Eq.~(\ref{change-var})) and the expansion (Eq.~(\ref{p-expan})), the boundary conditions for Eq.~(\ref{eq-p0-p1}) read as: $\bar{p}_0(\bar{r},-\bar{L}) = \bar{L}$, $\bar{p}_0(\bar{r},\bar{h}) = -\bar{h}_{J}$ and $\bar{p}_1(\bar{r},-\bar{L}) = \bar{p}_1(\bar{r},\bar{h}) = 0$. Eq.~(\ref{eq-p0-p1}) together with their boundary conditions are easily solved to obtain the pressure $p$ which, once substituted in Eq.~(\ref{vr-vz}), gives the velocity field, see SI Appendix. This velocity field evaluated at $z=h(r,t)$ is then used in Eq.~(\ref{h-eq}) to obtain the spatio-temporal evolution of the air-liquid interface
{\small
\begin{subequations}
\begin{align}
\label{eq-H-adim}
&\frac{\partial H}{\partial \bar{t}} = 1 - \frac{H}{\overline{V}[H + \bar{h}_J + \bar{L}]}+ \frac{\bar{\delta}^2}{3\overline{V}} \frac{\bar{L}^3[\bar{L}+\bar{h}_J]}{[H+\bar{h}_J + \bar{L}]^3}\nabla_{\bar{r}}^2 H, \\
&H=\frac{h-h_J}{L_0}, \quad \overline{V} = \frac{V}{V_{\parallel}}, \quad \bar{\delta}^2 = \frac{\delta^2 k_{\perp}}{k_{\parallel}}, \quad V_{\parallel}=\frac{k_{\parallel} \rho g}{\mu},
\end{align}
\end{subequations}
}
where $H$ is the dimensionless height of the interface with respect to its initial (static) position, $\bar{L}=L/L_0 = 1-\bar{t}$ the relative variation of the immersion depth, and $H + \bar{h}_J + \bar{L} = (h + L)/L_0$ is the dimensionless (time-varying) wet length of the brush.

As expected, Eq.~(\ref{eq-H-adim}) involves the 3 dimensionless parameters $\overline{V}$, $\overline{\delta}$ and $\overline{h}_j$ identified earlier. For $\overline{V}\rightarrow \infty$ only the first term in Eq.~(\ref{eq-H-adim}) remains and $H(\overline{t})=\overline{t}$ (i.e the fluid moves upwards with the brush). For finite $\overline{V}$, there is a competition between this upwards motion and the fluid flowing out of the brush according to the last two terms in Eq.~(\ref{eq-H-adim}). The first of these two terms describes the vertical flow and involves the ratio between the height of the fluid on which gravity forces act and the wet length of the brush along which there is viscous friction. The last term results from the flow in the radial direction and is proportional to $\overline{\delta}$, the vertical to horizontal flow rate ratio identified earlier.

Eq.~(\ref{eq-H-adim}) requires an initial condition and two boundary conditions to be solved in the domain $0 \le (\bar{r},\bar{t}) \le 1$:
\begin{equation}
\label{BC}
H(\bar{r},0)=0, \quad \partial_{\bar{r}} H(\bar{r},\bar{t})|_{\bar{r} = 0} = H(1,\bar{t}) =0.
\end{equation}
The initial condition comes from the definition of $H$, the first boundary conditions stems from symmetry and the second one from the assumption that the pressure in the bath is hydrostatic for $r\ge D/2$. In the limit of an infinitely wide brush, $L_0/D \to 0$, the aspect ratio $\delta$ vanishes and the flow is everywhere uni-directional along the $z$-axis so that Eq.~(\ref{eq-H-adim}) becomes an ODE describing the motion of a flat horizontal interface. Due to the perturbative scheme used to derive Eq.~(\ref{eq-H-adim}), it is expected to hold only when $\delta \ll 1$. We show below that a good agreement with experiments is actually obtained up to $\delta$ of order 1.

Figure~\ref{fig02}B shows the spatio-temporal evolution of the air-liquid interface, $H(\bar{r},\bar{t})$, obtained by solving numerically Eq.~(\ref{eq-H-adim}) for some typical values of $\bar{\delta}$ and $\bar{h}_J$, when a brush is removed from a liquid bath at two different retraction speeds. The height of the interface, measured along the central axis of the brush, $H_0(\bar{t}) \equiv H(0,\bar{t})$, reaches a maximum value before the end of retraction, occurring at $\bar{t}=1$, for low retraction speeds whereas it grows monotonically at large enough speeds. The transition between these two types of behaviors occurs at a speed $\overline{V}_{T}\approx 1/(1+2.63\, \bar{h}_J^{6/5})$ (see SI appendix). 

Once $H$ is obtained numerically, the pressure and velocity fields can be computed from their analytical expressions given in SI Appendix. Fig.~\ref{fig02}C and D show, respectively, the pressure and velocity fields at $\bar{t}=0.66$ for the evolution of $H$ shown in the top panel of Fig.~\ref{fig02}B. The pressure gradient is essentially constant along $z$ and vanishing along $r$ almost everywhere within the brush except near the rim ($\bar{r}=1$) where it is steeper in both directions leading to a larger magnitude of the velocity. The velocity field is thus mainly unidirectional near the central axis and bidirectional near the rim where the liquid escapes radially from the brush. For that value of $\bar{t}$, the interface is moving downward ($H_0^m$ is reached at $\bar{t}=0.56$). The velocity is hence oriented toward the bath except near the rim where the liquid is moving upward and also radially near the bath, which results in a region where the sign of $v_z$ changes ($\bar{r} \simeq 0.7$). 

An important test of the model is the rationalization of the complex evolution of $H_0^m$ shown in Fig.~\ref{fig01}F, where the largest values of $H_0^m$ are not necessarily reached for the largest retraction speeds or immersion depths according to the porosity of the brushes. This complexity is related to the existence of 3 dimensionless groups of parameters governing the dynamics: $\overline{V}$, $\bar{h}_J$ and $\bar{\delta}$. To get insight into the variation of $H_0^m$, we show in Fig.~\ref{fig02}E its variation as a function of the retraction speed for several values of $\bar{\delta}$ and $\bar{h}_J$. When $\overline{V} \lesssim \overline{V}_{T}$, i.e. when $H_0^m$ is reached at $\bar{t} < 1$, $H_0^m$ evolves linearly with $\overline{V}$ and writes as $H_0^m = C(\bar{h}_J,\bar{\delta}) \overline{V}$. The evolution of $C(\bar{h}_J,\bar{\delta})$ with $\bar{h}_J$ is given in Fig.~\ref{fig02}F for several values of $\bar{\delta}$. As explained, the overall variation of $H_0^m$ involves the 3 dimensionless control parameters and reads as
\begin{equation}
\label{Deltah0m}
H_0^m \simeq [\bar{h}_J+\alpha(\bar{\delta})]\, \overline{V}, \quad \overline{V} \lesssim \overline{V}_{T},
\end{equation}
where $\alpha$ is shown in the inset of Fig.~\ref{fig02}F and is well approximated by $\alpha^2(x)\approx \tanh[1/(2 x)^2]$. 

\begin{figure}[!t]
\centering
\includegraphics[width=0.8\columnwidth]{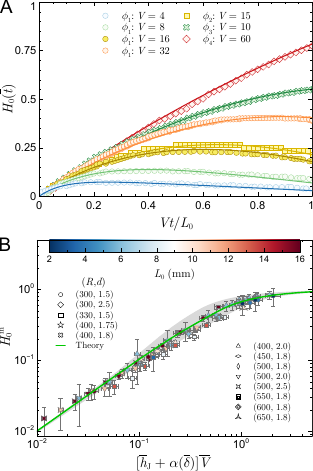}
\caption{(A) Comparison between some temporal variations of $H_0(\bar{t})$ measured experimentally (symbols) and computed from Eq.~(\ref{eq-H-adim}) (solid curves) for 4 different brushes: $\phi_1 = 0.77$ ($L_0 = 6$ mm), $\phi_2=0.81$ ($L_0 = 6$ mm), $\phi_3= 0.53$ ($L_0 = 10$ mm), $\phi_4 = 0.85$ ($L_0 = 2$ mm). The retraction speed $V$ is given in mm/min and $\mu = 0.97$ Pa s. The corresponding dimensionless parameters vary in the range: $0.06 \le \overline{V} \le 0.91$, $0.15 \le \bar{\delta} \le 0.78$ ($0.22 \le \delta \le 1.27$) and $0.44 \le \bar{h}_J \le 1.27$. (B) Evolution of $H_0^m$ measured experimentally (symbols) as a function of a rescaled retraction speed $[\bar{h}_J + \alpha(\bar{\delta})]\overline{V}$, where $\alpha$ is shown in the inset of Fig.~\ref{fig02}E, for 13 different brushes where $2 \le V \le 150$ mm/min, $15 \le D \le 21.2$ mm and $\mu=0.97$ Pa s. The number of pillars varies between 60 and 163. The grey area shows the region spanned by the theory when $\bar{\delta}$ and $\bar{h}_J$ are varies within the experimental range ($0.15 \le \bar{\delta} \le 1.30$, $0.05 \le \bar{h}_J \le 1.27$). The green curve is computed using their average value: $\bar{\delta}=0.73$ and $\bar{h}_J=0.35$.}
\label{fig03}
\end{figure}

\begin{figure*}[!t]
\centering
\includegraphics[width=\textwidth]{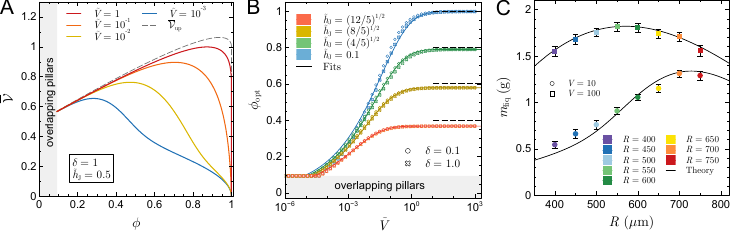}
\caption{(A) Evolution of the rescaled volume captured by a brush, $\overline{\mathcal{V}}$, defined in Eq.~(\ref{Volume}), as a function of the porosity, $\phi$, for $\delta =1$, $\tilde{h}_J=0.5$ and various values of $\widetilde{V}$ (see Eq.~(\ref{eq-p-adim}) and Eq.~(\ref{new-parameters}) for the definition of these parameters). The dashed curve shows the limiting value $\overline{\mathcal{V}}_{\text{up}} = \phi + \tilde{h}_{J} (1-\phi)^{1/2}$ reached at large retraction speed when the column of liquid initially inside the brush is entirely pull out the bath. (B) Evolution of the optimal porosity, $\phi_{\text{opt}}$, corresponding to the maximum of $\overline{\mathcal{V}}$ shown in panel (A), as a function of $\widetilde{V}$ for several values of $\delta$ and $\tilde{h}_J$. The horizontal dashed are analytical estimations, $\phi_{\text{opt}} = 1- \tilde{h}_{J}^2/4$, obtained by using $\overline{\mathcal{V}}_{\text{up}}$ for the volume of liquid captured. (C) Measured mass of liquid, $m_{\text{liq}}$, captured at the end of the retraction process by brushes with various pillar radii $R$ (expressed in $\mu$m) for two retraction speeds (expressed in mm/min) and $d=1.8$ mm, $L_0 = 10$ mm, $\mu = 0.97$ Pa s. The solid curves correspond to the mass computed with the theory. $D=15.8$ mm and $D=16.9$ mm were used in the theory for $V = 10$ mm/min and $V = 100$ mm/min, respectively.}
\label{fig04}
\end{figure*}

\textbf{Comparison with experiments.} Figure~\ref{fig03}A shows a comparison between some typical temporal evolution of $H_0$ obtained experimentally with various brushes and the corresponding theoretical evolution obtained by solving numerically Eq.~(\ref{eq-H-adim}). A good agreement between theory and experiments is observed for various retraction speeds and immersion depths even when $\delta$ is of order 1 and without any fitting parameter. Fig.~\ref{fig03}B shows a good collapse of the experimental data onto the theoretical prediction when the raw data for $h_0^m-h_J$ reported in Fig.~\ref{fig01}F are rescaled by $L_0$ and plotted as a function of the new dimensionless group identified in~Eq.~(\ref{Deltah0m}). The grey area in Fig.~\ref{fig03}B, where essentially all the data lie, highlights the region spanned by the theoretical variation of $H_0^m$ when the parameters $\bar{h}_J$ and $\bar{\delta}$ are varied within their experimental range. The solid green curve shows the theoretical evolution of $H_0^m$ when the experimental average value of $\bar{h}_J$ and $\bar{\delta}$ is used.

\textbf{Optimal brush design.} Having validated the model through a confrontation with numerous experimental data, we now focus our attention on the volume $\mathcal{V}$ of liquid captured at the end of retraction ($\bar{t} = 1$), which reads as:
\begin{equation}
\label{Volume}
\overline{\mathcal{V}} = \frac{\mathcal{V}}{\mathcal{V}_I} = 2 \phi \left[\int_0^1 H(\bar{r},1)\, \bar{r}\, d\bar{r} + \frac{\bar{h}_J}{2}\right],
\end{equation}
where $\mathcal{V} = 2\pi \phi \int_0^{D/2} h(r,L_0/V)\, r\, dr$ and $\mathcal{V}_I = \pi D^2 L_0/4$. $\phi\mathcal{V}_I$ is thus the volume of liquid initially inside the brush below the bath level before retraction starts. Since $H(\bar{r},1)$ can never be larger than 1, Eq.~(\ref{Volume}) implies that $\overline{\mathcal{V}} \le \overline{\mathcal{V}}_{\text{up}} = \phi (1 + \bar{h}_J)$.
In experiments with a given brush and a given liquid, only the retraction speed $V$ and the immersion depth $L_0$ can be varied. Figure~\ref{fig01}D and E show that the height reached by the interface at the end of retraction ($\bar{t} = 1$), and thus the volume $\mathcal{V}$, grows monotonically with $V$ and $L_0$ respectively for a given $\phi$. There is thus no optimal retraction speed or immersion depth for a given brush capturing a given liquid. However, for a given liquid and given $V$ and $L_0$, there exists an optimal brush maximizing $\mathcal{V}$. Indeed, increasing the porosity $\phi$ increases the volume available for the liquid inside the brush but also increases the permeabilities [Fig.~\ref{fig02}A] which reduces the height reached by the interface. The interplay between these two antagonistic contributions leads to a non-monotonic variation of $\mathcal{V}$ with $\phi$.

To determine the optimal porosity, we rewrite the parameters in Eq.~(\ref{eq-H-adim}) as follows to make explicit the dependence on $\phi$
\begin{equation}
\label{new-parameters}
\overline{V} = \frac{\widetilde{V}}{F_{\parallel}(\phi)}, \quad \frac{\bar{\delta}^2}{\overline{V}} = \frac{\delta^2}{\widetilde{V}} F_{\perp}(\phi), \quad \bar{h}_J = \tilde{h}_J \frac{\sqrt{1-\phi}}{\phi},
\end{equation}
where $\widetilde{V} = \mu V/(\rho g d^2)$, $\tilde{h}_J = 2 \sqrt{2\pi}\ell_c^2 \cos \theta_Y /(3^{1/4} d L_0)$, and the expressions of $F_{\parallel}$ and $F_{\perp}$ given in Fig.~\ref{fig02}A. $\tilde{h}_J$ is obtained from Eq.~(\ref{jurin-hex}) with $R$ expressed as a function of $d$ and $\phi$. Consequently, $H$ and thus $\overline{\mathcal{V}}$ depend only on $\phi$ when $\widetilde{V}$, $\delta$ and $\tilde{h}_J$ are fixed. We consider here that a variation of $\phi$ is due to a change of $R$ with $d$ fixed. By expressing $d$ as a function of $R$ and $\phi$, we could alternatively consider that the variation of $\phi$ results from a change of $d$ with $R$ fixed with similar conclusions with respect to the existence of an optimal porosity. 

Figure~\ref{fig04}A shows the non-monotonic evolution of the rescaled volume $\overline{\mathcal{V}}$ with $\phi$ for various values of $\widetilde{V}$ with $\delta=1$ and $\tilde{h}_J = 1/2$. This behavior can be explained as follows. For a given value of $\widetilde{V}$, Eq.~(\ref{new-parameters}) together with Fig.~\ref{fig01}B show that the rescaled retraction speed $\overline{V}$ is large when $\phi$ is small. In this limit, Eq.~(\ref{eq-H-adim}) shows that $H(\bar{r},1) \simeq 1$ so that, according to Eq.~(\ref{Volume}), $\overline{\mathcal{V}} \simeq \overline{\mathcal{V}}_{\text{up}}=\phi + \tilde{h}_J \sqrt{1-\phi}$. The amount of liquid captured thus grows with $\phi$ when the latter is small enough. In contrast, when $\phi\to 1$, there is no capillary rise or viscous entrainment and $\overline{\mathcal{V}}$ vanishes. Therefore, $\overline{\mathcal{V}}$ necessarily features a maximum value at some intermediate value of $\phi$. 

Figure~\ref{fig04}B shows the variations of the optimal porosity, $\phi_{\text{opt}}$, as a function of $\widetilde{V}$ for several values of $\delta$ and $\tilde{h}_J$. At large retraction speed, $\phi_{\text{opt}}$ saturates to a constant value which can be estimated using the simple expression of the upper limit $\overline{\mathcal{V}}_{\text{up}}$ which features a maximum for $\phi = 1-\tilde{h}_J^2/4$. This estimation is shown as horizontal dashed lines in Fig.~\ref{fig04}B and its difference with the numerical results is due to the flatness of $\overline{\mathcal{V}}$ around its maximum as seen in Fig.~\ref{fig04}A. For example, for $\tilde{h}_{J} = (12/5)^{1/2}$, the relative error on the limiting value of $\phi_{\text{opt}}$ at large $\widetilde{V}$ is about 9\% but the relative error on volume captured $\overline{\mathcal{V}}$ is only about 0.025\%. Neglecting the small influence of $\delta$, the evolution of $\phi_{\text{opt}}$ is rather well described by the following expression
\begin{subequations}
\begin{align}
\phi_{\text{opt}} &\approx A(\tilde{h}_J) \tanh\left(B(\tilde{h}_J)\,\widetilde{V}^{1/4} \right), \\
A &= 1-0.263\, \tilde{h}_{J}^2, \quad B = 1.7 + 0.041 \, e^{2.5 \, \tilde{h}_{J}}.
\end{align}
\end{subequations}
The optimal porosity varies thus significantly when $\widetilde{V} \lesssim 1$ and saturates to a constant value when $\widetilde{V} \gtrsim 1$.

The existence of an optimal porosity has been verified experimentally. For this purpose, various brushes with a fixed number of pillars ($N_p = 73$), a fixed distance between their centers ($d=1.8$ mm) and various pillar radii ($400 \le R \le 750$ $\mu$m) were immersed at a given depth ($L_0 = 10$ mm) and removed at two retraction speeds ($V = 10$ mm/min and $V=100$ mm/min) (Movie S4). The mass of liquid transported by these brushes when they are displaced by a distance $L_0$ has been measured by the traction device and is reported in Fig.~\ref{fig04}C. This mass is maximum for $R \simeq 700-750$ $\mu$m when $V = 10$ mm/min and $R \simeq 550-600$ $\mu$m when $V = 100$ mm/min. These measurements are well described by the theory which predicts an optimal value $R=729$ and $R=571$ $\mu$m for $V = 10$ mm/min and $V = 100$ mm/min, respectively. 

\textbf{Conclusion.} In summary, we have studied experimentally and theoretically the liquid captured when a rigid brush-like structure is withdrawn from a liquid bath. We experimentally observe how the amount of liquid transported by viscous entrainment depends on the brush porosity, the retraction speed and the immersion depth. This intricate dependence stems from the two-dimensional nature of the drainage flow and on its intricate time evolution with the brush withdrawal. We developed a theoretical analysis in which this three dimensional flow is solved as a perturbation of the main gravity-driven vertical flow. This analysis yields a partial differential equation that describes the spatio-temporal evolution of the air-liquid interface within the brush, and hence the evolution of the volume of liquid entrained by the brush at any time. The obtained equation depends on three dimensionless parameters that reveal the physical ingredients governing the capture of a fluid by viscous entrainment.

Our model shows excellent agreement with our experimental data. In particular, the (quasi) master curve obtained for the maximum height reached by the liquid within the brush during the retraction provides a universal law to predict the maximum amount of liquid that can be captured by viscous entrainment. Thanks to the understanding offered by our model, we have also determined under which conditions the amount of liquid captured at the end of the retraction is maximum. We showed that, for a given velocity and immersion depth, an optimal brush porosity exists. We computed it as a function of the system parameters and verified it experimentally. 

The approach we developed here can be extended, in a straightforward manner, to other fiber arrangements using the appropriate longitudinal and transverse permeability coefficients. A natural prolongation of this work would be the study of soft brushes and the impact of the deformability of the structures on the liquid capture. The role of capillarity in the pinch-off~\cite{Eggers1993} of the liquid bridge between the brush and the bath as the structure is fully removed and in the amount of liquid remaining in the brush when the drainage is completed are also yet to be rationalized. As a whole, this work provides experimental and theoretical advances to the physics of anisotropic porous media. In particular, the identification of optimal brushes for fluid capture has the potential to influence engineering applications in liquid manipulation and transfer.

\begin{acknowledgments}
We acknowledge support by the Fund for Scientific Research (F.R.S.-FNRS) under the Research Grants No. T.0025.19 (PDR ``ElastoCap'') and No. J.0017.21 (CDR ``FASTER'') and by the Federation Wallonia-Brussels (FWB) (Concerted Research Actions ``Capture''). This project has received funding from the European Union's Horizon 2020 research and innovation programme under the Marie Sklodowska-Curie Grant Agreements No. 101027862 and No. 101102728. We thank the Micro-milli service platform (ULB) for the access to their experimental facilities and Adam Chafaï for his help on manufacturing the brushes.
\end{acknowledgments}

{\small
\textbf{Methods: Brushes Preparation and experimental apparatus.} The brushes were printed with a PolyJet 3D printer Eden260 from Stratasys, using VeroWhite or VeroClear resins. Holders for the brushes were 3D printed using a FDM printer Ultimaker S5 and then glued to the brush with superglue. A vessel was filled with silicon oil (V100 or V1000) from Sigma-Aldrich. The brushes with their holder where attached to a gripper on a traction device (ZwickiLine Z0.5 from ZwickRoell) and dipped into the fluid at a given depth $L_0$. When the capillary rise had reached its final height and that no further flow was present within the structure, the brush was removed at a constant speed $V$ from the bath. The removal was recorded from the side with LED backlighting, using a Basler CMOS camera with a frame rate adjusted to the withdrawal speed. The height $h_0$ of the interface along the brush central axis was measured as a function of time using standard image analysis techniques using Python routines and ImageJ. The force acting on the brush during retraction was also recorded using a 10N force sensor from ZwickRoell.
}

\bibliography{brush.bib}

\clearpage
\onecolumngrid

\setcounter{equation}{0}

{\Large \textbf{Supplemental Material}}

\section{Jurin's height}

When a solid with small interstices compared to the capillary length is put into contact with a liquid bath, the liquid rises inside the pores up to the so-called Jurin's height minimizing the surface energy, $U_S$, and the work of the weight of the liquid rising in the pores, $U_G$:
\begin{equation}
\label{ener-jurin-gen}
U = U_S + U_G = \gamma A_p + \gamma_{LS} A_{LS} + \gamma_{SV} A_{SV} + \rho g \int_V z\, dV,
\end{equation}
where $\gamma$, $\gamma_{LS}$, $\gamma_{SV}$ are, respectively, the air-liquid, liquid-solid, solid-air surface energies, $A_p$, $A_{LS}$, $A_{SV}$ are, respectively, the area of the air-liquid, liquid-solid, solid-air interfaces, $\rho$ is the liquid density and $V$ the volume of the liquid rising in the pores. Since the solid is either in contact with the liquid or with air, we have $A_{LS} + A_{SV} = A_T$, where $A_T$ is the total area of the solid. In addition, considering a solid whose geometry is invariant along the vertical $z$-direction, we have $A_{LS} = z\, \ell_{LS}$, where $\ell_{LS}$ is the length of the liquid-solid contact line in a given horizontal plane, and $V = z\, A_p$ where $0\le z\le h_J$. Finally, the Young-Laplace-Dupret law gives a relationship between the surface energies: $\gamma_{SV} - \gamma_{LS} = \gamma \cos \theta_Y$, where $\theta_Y$ is the Young contact angle. Using these relations, Eq.~(\ref{ener-jurin-gen}) evaluated at $z=h_J$ becomes
\begin{equation}
\label{ener-jurin}
U(h_J) = \gamma A_p - \gamma \cos \theta_Y \ell_{LS} h_J + \gamma_{SV} A_{T} + \frac{\rho g}{2} A_p h_J^2.
\end{equation}
The minimum of the energy is obtained from $dU/dh_J=0$ and yields the Jurin's height
\begin{equation}
\label{jurin-gen}
h_J = \frac{\gamma}{\rho g} \frac{\ell_{LS}}{A_p} \cos \theta_Y = \ell_c^2 \frac{\ell_{LS}}{A_p} \cos \theta_Y,
\end{equation}
where we have introduced the capillary length $\ell_c$. For a capillary tube of circular cross-section of radius $R$, we have $\ell_{LS}=2\pi R$ and $A_p = \pi R^2$, so that $h_J = 2\ell_c^2 R^{-1} \cos \theta_Y$ as it should~\cite{PGG2004}.

\begin{figure*}[!t]
\centering
\includegraphics[width=0.72\textwidth]{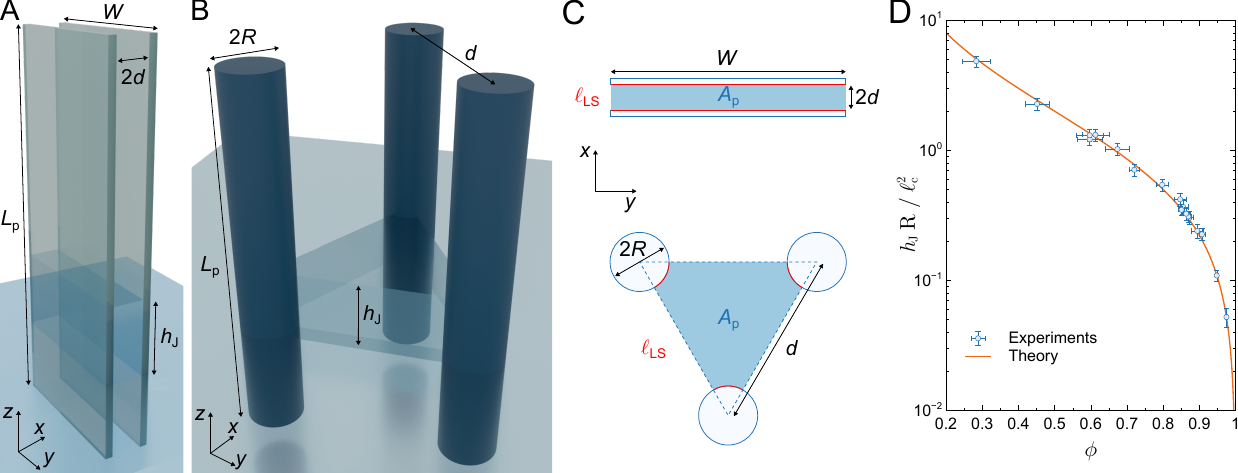}
\caption{\textbf{Schematics of the systems.} (A) Two parallel plates of width $W$ and separated by a distance $2d$ partially immersed in a liquid bath. The liquid raises up to $z=h_J$ in between the plates. (B) Unit cell of an equilateral triangular array of cylinders of radius $R$ and separated by a distance $d$. The liquid raises up to $z=h_J$ in between the cylinders. (C) Horizontal cross-sections of the systems shown in panels (A) and (B) highlighting the pore area $A_p$ and the length of the solid-liquid interface $\ell_{LS}$. (D) Experimental measurements of the Jurin's height in an equilateral triangular array of cylinders with various porosities together with the theoretical prediction (\ref{Jurin-th}).}
\label{FigS-Jurin}
\end{figure*}

For two parallel plates of width $W$ and separated by a distance $2d$, see Fig.~\ref{FigS-Jurin}A and C, we have $\ell_{LS} = 2W$ and $A_p = 2d W$ so that
\begin{equation}
h_J = \frac{\ell_c^2}{d} \cos \theta_Y.
\end{equation}
For an equilateral triangular array of cylinders of radius $R$ and separated by a distance $d$, see Fig.~\ref{FigS-Jurin}B and C, we have $\ell_{LS}=N\pi R$ and $A_p = N[\sqrt{3}d^2-2\pi R^2]/4$, where $N$ is the number of unit cells in the solid. Eq.~(\ref{jurin-gen}) becomes
\begin{equation}
\label{Jurin-th}
h_J = \frac{2\ell_c^2}{R}\left[\frac{1-\phi}{\phi}\right]\cos \theta_Y, \quad \phi = 1-\frac{2\pi}{\sqrt{3}}\frac{R^2}{d^2},
\end{equation}
where $1-\pi/(2\sqrt{3})\le \phi \le 1$ is the porosity of the solid~\cite{Princen1969}. This theoretical expression compares well with experimental data reported in Fig.~\ref{FigS-Jurin}D.

Note that for a square array of cylinders of radius $R$ and separated by a distance $d$, we have the same expression for $h_J$ in terms of the porosity. Indeed, in this case, we have $\ell_{LS} = 2N\pi R$ and $A_p = N[d^2-\pi R^2]$, where $N$ is the number of unit cells in the solid. Eq.~(\ref{jurin-gen}) becomes
\begin{equation}
h_J = \frac{2\ell_c^2}{R}\left[\frac{1-\phi}{\phi}\right]\cos \theta_Y, \quad \phi = 1-\frac{\pi R^2}{d^2},
\end{equation}
where $1-\pi/4\le \phi \le 1$ is the porosity of the solid~\cite{Princen1969}.

\section{Permeability of an equilateral triangular array of pillars}

When the porosity is sufficiently close to 1, the asymptotic expressions for the permeabilities of an equilateral triangular array of pillars are given by 
\begin{equation}
\label{k-para-asymp}
\frac{k_{\parallel}}{R^2} = \frac{1}{4(1-\phi)}\left[-\ln(1-\phi) - K + 2(1-\phi)-\frac{(1-\phi)^2}{2}\right], \quad \frac{k_{\perp}}{k_{\parallel}} \simeq 1/2,
\end{equation}
where $K= 1.498$~\cite{Drummond1984,Jackson1986}. Note that this is the same expression for a square arrays of cylinders with $K = 1.476$~\cite{Drummond1984,Jackson1986}. Expansions at low porosity exist also for $k_{\parallel}$ but not for $k_{\perp}$~\cite{Drummond1984}. For this reason, we used numerical simulations performed with the COMSOL Multiphysics\textsuperscript\textregistered software to compute $k_{\parallel}$ and $k_{\perp}$ for all ranges of $\phi$ and fitted the numerical data as explained below.

\begin{figure*}[!b]
\centering
\includegraphics[width=\textwidth]{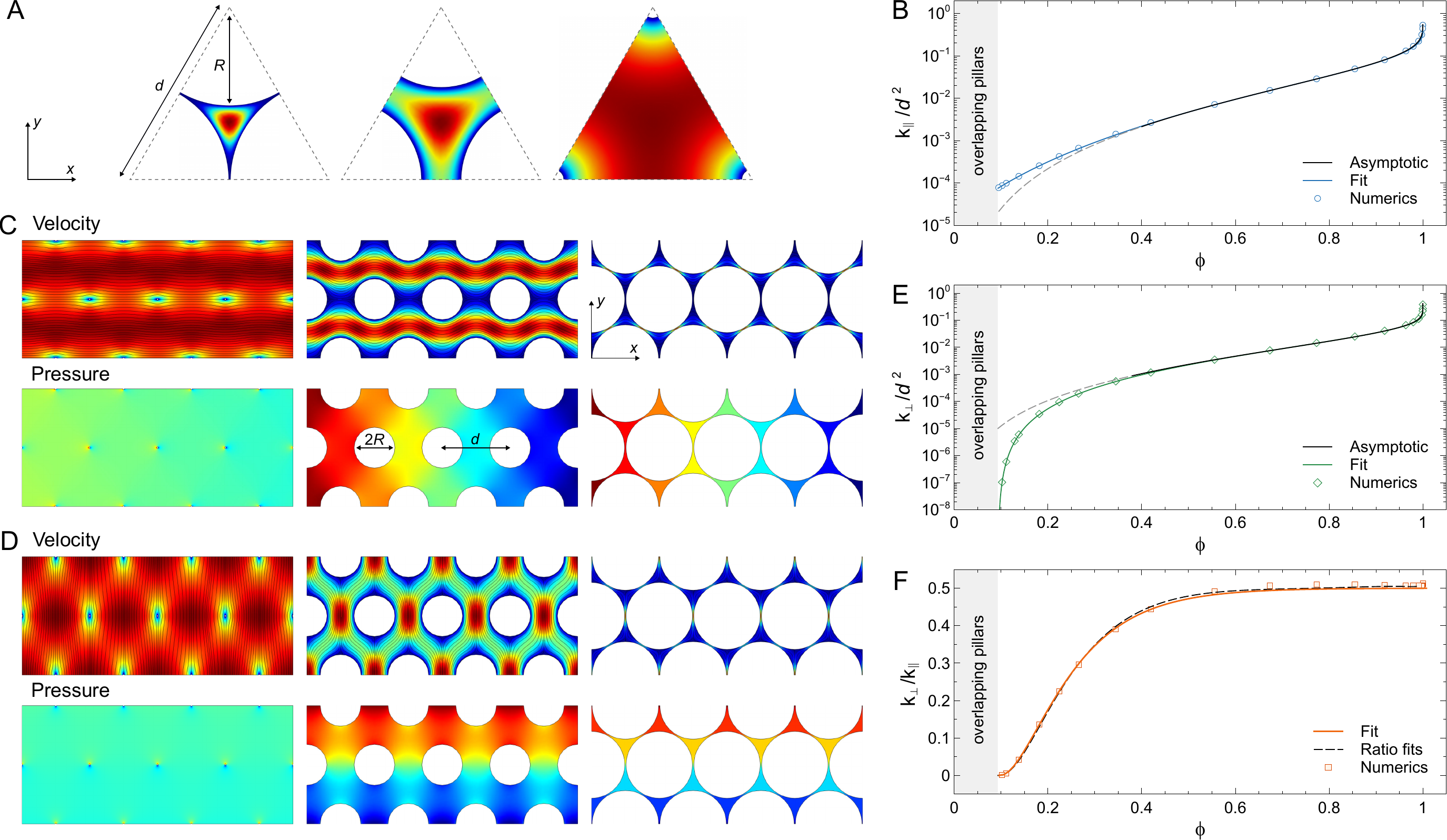}
\caption{\textbf{Permeabilities of an equilateral triangular of cylinders.} (A) Magnitude of the velocity field of a longitudinal flow along the $z$-axis for three value of $R$ at a given $d$. (B) Evolution of $k_{\parallel}/d^2$ as a function of the porosity $\phi$ obtained numerically together with the asymptotic relation (\ref{k-para-asymp}) and the fit (\ref{fit-k-para}). (C) Magnitudes of the velocity and pressure fields for a transverse flow along the $x$-axis for three value of $R$ at a given $d$. (D) Same as panel (C) for a transverse flow along the $y$-axis. Both cases lead to the same value of the permeability. (E) Evolution of $k_{\perp}/d^2$ as a function of the porosity $\phi$ together with the asymptotic relation (\ref{k-para-asymp}) and the fit (\ref{fit-k-perp}). (F) Evolution of the ratio $k_{\perp}/k_{\parallel}$ as a function of $\phi$ together with the ratio of the fits (\ref{fit-k-perp}) and (\ref{fit-k-para}) and a simpler fit (\ref{fit-k-perp-simple}).}
\label{FigS-perm}
\end{figure*}

To compute the longitudinal permeability $k_{\parallel}$, an equilateral triangular domain is used in a $(x,y)$ plane where the length of each side is equal to $d$ and where cylinders of radius $R$ are centered at each vertex, see Fig.~\ref{FigS-perm}A. This domain extends along the $z$-axis over a length $L_z$. Non slip and symmetric boundary conditions are imposed on the cylinder walls and at the interstices respectively. A difference of pressure $\Delta p$ is applied at the two extremities of the system along the $z$-axis and the resulting flow rate $Q$ is computed. The longitudinal permeability is then given by
\begin{equation}
k_{\parallel} = \frac{\mu L_z Q}{S_{xy} \Delta p}, \quad S_{xy} = \frac{\sqrt{3}\, d^2}{4}.
\end{equation}
Fig.~\ref{FigS-perm}B shows that the asymptotic expression (\ref{k-para-asymp}) agrees well with the numerical data up to $\phi \simeq 0.4$. This figure shows also that a good description of the numerical data over the entire range of $\phi$ is obtained with the following fit
\begin{equation}
\label{fit-k-para}
\frac{k_{\parallel}}{d^2} \approx \frac{1}{4}\left[1-\frac{\sqrt{3}\, R}{d}\right]^{4.6}\left[1-\frac{\sqrt{3}}{8\pi}\left(\ln(1-\phi)+K-35(1-\phi)\right)\right],
\end{equation}
where the first factor has been added to obtain $k_{\parallel} = 0$ when $R/d = 1/\sqrt{3} \simeq 0.577$ since, in this case, there is no interstice between the cylinders and there is no flow. Note that this configuration is only possible with overlapping cylinders. With non-overlapping cylinders, the maximal value of $R/d$ is $1/2$ so that the minimal value of $\phi$ is $1-\pi/(2\sqrt{3}) \simeq 0.093$. Note also that $d^2$ is chosen as the length scale for $k_{\parallel}$ in Eq.~(\ref{fit-k-para}) instead of $R$ as in Eq.~(\ref{k-para-asymp}).

To compute the transverse permeability $k_{\perp}$, a two-dimensional domain of size $L_x =4d$ and $L_y=\sqrt{3}d$ with periodic boundary conditions along $x$ and $y$ is used and non-slip boundary conditions on the cylinder walls. A difference of pressure $\Delta p $ is applied at two extremities along the $x$ or the $y$-axis, see Fig.~\ref{FigS-perm}C,D, and the resulting flow rate $Q$ per unit length is computed. The permeability is then given by
\begin{equation}
k_x = \frac{\mu L_x Q}{L_{y} \Delta p}, \quad k_y = \frac{\mu L_y Q}{L_{x} \Delta p}.
\end{equation}
The numerical simulations give $k_x = k_y \equiv k_{\perp}$ as it should. Fig.~\ref{FigS-perm}E shows that the asymptotic expression (\ref{k-para-asymp}) agrees well with the numerical data up to $\phi \simeq 0.4$ as for $k_{\parallel}$. This figure shows also that a good description of the numerical data over the entire range of $\phi$ is obtained with the following fit
\begin{equation}
\label{fit-k-perp}
\frac{k_{\perp}}{d^2} \approx \frac{1}{12.4}\left[1-\frac{2R}{d}\right]^{2.5} \left[1-0.21(\ln(1-\phi)+K)\right],
\end{equation}
where the first factor has been added to obtain $k_{\perp} = 0$ when $R/d = 1/2$ because, in this case, there is no interstice between the cylinders for a transverse flow as the pillars are in self-contact. Again, $d^2$ is chosen as the length scale for $k_{\perp}$ in Eq.~(\ref{fit-k-perp}) instead of $R$ as in Eq.~(\ref{k-para-asymp}).

Fig.~\ref{FigS-perm}F shows the evolution of $k_{\perp}/k_{\parallel}$ as a function of $\phi$ together with the ratio of the fits (\ref{fit-k-perp}) and (\ref{fit-k-para}). The good agreement between the data and this ratio highlights the good quality of the fits. Nevertheless, the ratio of these fits yields a rather cumbersome expression. A good fit of the data can be obtained with a much simpler expression 
\begin{equation}
\label{fit-k-perp-simple}
\frac{k_{\perp}}{k_{\parallel}} \approx \frac{1}{2} \left(1-a\exp[b (1-\phi)]\right)^2, \quad a = 3.4\times 10^{-4}, \quad b = 8.86.
\end{equation}

\section{Main equations for a brush removed at constant speed from a bath}

\subsection{Equation for the interface} 

The velocity and pressure fields of the fluid inside a brush removed along the $z$-axis at speed $V$ from a liquid bath is given by Darcy's law in cylindrical coordinates 
\begin{equation}
\label{vr-vz-dim}
v_{r}(r,z) = -\frac{k_{\perp}(z)}{\mu} \frac{\partial p}{\partial r}, \quad
v_z(r,z) = V - \frac{k_{\parallel}}{\mu} \left(\frac{\partial p}{\partial z} + \rho g\right), \quad \frac{\partial^2 p}{\partial z^2} +\frac{k_{\perp}(z)}{k_{\parallel}}\nabla_r^2 p =0,
\end{equation}
where the equation for the pressure is obtained from mass conservation with $\nabla_r^2 p = r^{-1}\partial_r(r \partial_r p)$ and $z=0$ corresponds to the position of the air-liquid interface of the bath. We consider that the transverse permeability depends on $z$ because we assume that there is no flow along the transverse (horizontal) direction inside the fluid transported by the brush located above the level of the liquid bath. Indeed, this fluid region is surrounded by an air-liquid interface preventing any radial expansion during the brush withdrawal; the radial velocity should thus be small in this region. Therefore
\begin{equation}
\label{kpz}
k_{\perp}(z) = k_{\perp} \theta(-z),
\end{equation}
where $\theta$ is the Heaviside function. Note that using $k_{\perp}$ constant everywhere underestimates the height of the liquid column entrained by the brush compared to experiments since the fluid can flow horizontally everywhere. In addition, the resulting equation for the spatio-temporal evolution of the air-liquid interface depends only on two dimensionless parameters whereas experiments show that the dynamics is governed by at least 3 dimensionless groups. Using Eq.~(\ref{kpz}) addresses both of these issues but introduces a discontinuity in $v_r$ at $z=0$. This features mimics a transition along $z$ where $v_r$ decreases significantly over a short distance near $z=0$. We did not attempt to regularized the Heaviside function by using a smooth continuous function since this would introduce an additional parameter controlling the sharpness of the transition. 

Using the change of variables
\begin{equation}
\label{change-var-sup}
\bar{r} = 2r/D, \quad (\bar{z},\bar{h},\bar{L},\bar{h}_{J},\bar{t},\bar{p}) = (z,h,L,h_J,V t, p/\rho g)/L_0,
\end{equation}
the equation for the pressure becomes
\begin{equation}
\label{eq-p-adim-sup}
\frac{\partial^2 \bar{p}}{\partial \bar{z}^2}+ \frac{k_{\perp}(z)}{k_{\parallel}} \delta^2 \nabla_{\bar{r}}^2 \bar{p} =0, \quad \delta = \frac{2L_0}{D},
\end{equation}
where the aspect ratio $\delta$ is assumed to be small. Expanding the pressure up to order $\delta^2$, $\bar{p} = \bar{p}_0 + \delta^2\, \bar{p}_1$, 
Eq.~(\ref{eq-p-adim-sup}) becomes at orders $\delta^0$ and $\delta^2$
\begin{equation}
\label{eq-p0-p1-sup}
\frac{\partial^2 \bar{p}_0}{\partial \bar{z}^2}=0, \quad \frac{\partial^2 \bar{p}_1}{\partial \bar{z}^2}=- \frac{k_{\perp}(z)}{k_{\parallel}}\nabla_{\bar{r}}^2 \bar{p}_0.
\end{equation}
These equations are solved with the boundary conditions (see main text)
\begin{equation}
\bar{p}_0(\bar{r},-\bar{L}) = \bar{L}, \quad \bar{p}_0(\bar{r},\bar{h}) = -\bar{h}_{J}, \quad \bar{p}_1(\bar{r},-\bar{L}) = 0, \quad \bar{p}_1(\bar{r},\bar{h}) = 0.
\end{equation}
The solutions are
\begin{subequations}
\label{p0-p1}
\begin{align}
\bar{p}_0(\bar{r},\bar{z}) &= (\mathcal{P}(\bar{r})-1)\, \bar{z} + \bar{L} \mathcal{P}(\bar{r}), \quad \mathcal{P} = \frac{\bar{h}(\bar{r})-\bar{h}_{J}}{\bar{h}(\bar{r})+\bar{L}}, \\
\bar{p}_1(\bar{r},\bar{z}) &= \frac{k_{\perp}}{6k_{\parallel}}\left[-\bar{z}^2 \left(\bar{z} + 3 \bar{L} \right) \theta(-z)+  \frac{2\bar{L}^3 (\bar{h}(\bar{r})-\bar{z})}{\bar{L}+\bar{h}(\bar{r})}\right]\nabla_{\bar{r}}^2 \mathcal{P}(\bar{r}).
\end{align}
\end{subequations}
Note that the solutions are obtained at a given arbitrary time and the time dependence of the various functions is not indicated. Using the changes of variables (\ref{change-var-sup}), Eq.~(\ref{vr-vz-dim}) for the velocities and Eq.~(4) of the main text for the spatio-temporal evolution of the interface become
\begin{subequations}
\begin{align}
\label{vr-vz2}
v_{r}(\bar{r},\bar{z}) &= -\delta\, \frac{k_{\perp}(\bar{z})}{k_{\parallel}} V_{\parallel} \frac{\partial \bar{p}}{\partial \bar{r}}, \quad
v_z(\bar{r},\bar{z}) = V - V_{\parallel} \left(\frac{\partial \bar{p}}{\partial \bar{z}} + 1\right), \quad V_{\parallel}=\frac{k_{\parallel} \rho g}{\mu}, \\
\label{h-eq-adim}
\frac{\partial \bar{h}(\bar{r},\bar{t})}{\partial \bar{t}} &= \frac{v_z(\bar{r},\bar{h})}{V} - \delta \frac{v_r(\bar{r},\bar{h})}{V} \frac{\partial \bar{h}(\bar{r},\bar{t})}{\partial \bar{r}} = \frac{v_z(\bar{r},\bar{h})}{V},
\end{align}
\end{subequations}
where we used $v_r(\bar{r},\bar{h})=0$ because $k_{\perp}(\bar{h})=0$ since $\bar{h}>0$, see Eqs.~(\ref{vr-vz-dim}) and (\ref{kpz}). Substituting Eq.~(\ref{vr-vz2}) into Eq.~(\ref{h-eq-adim}), we get
\begin{equation}
\label{h-eq-adim2}
\frac{\partial \bar{h}}{\partial \bar{t}} = 1 - \frac{1}{\overline{V}} \left(\frac{\partial \bar{p}}{\partial \bar{z}} + 1\right)_{\bar{z} = \bar{h}}= 1 - \frac{1}{\overline{V}} \left(\frac{\partial \bar{p_0}}{\partial \bar{z}} + \delta^2 \frac{\partial \bar{p_1}}{\partial \bar{z}} + 1\right)_{\bar{z} = \bar{h}}, \quad \text{with} \quad \overline{V} = \frac{V}{V_{\parallel}},
\end{equation}
where we used the expansion $\bar{p} = \bar{p}_0 + \delta^2\, \bar{p}_1$. Using Eq.~(\ref{p0-p1}), we have
\begin{equation}
\left. \frac{\partial \bar{p}_0}{\partial \bar{z}}\right|_{\bar{z} = \bar{h}} = \mathcal{P}(\bar{r})-1, \quad \left. \frac{\partial \bar{p}_1}{\partial \bar{z}}\right|_{\bar{z} = \bar{h}} = -\frac{k_{\perp}}{3k_{\parallel}}\frac{\bar{L}^3 }{\bar{L}+\bar{h}(\bar{r})}\nabla_{\bar{r}}^2 \mathcal{P}(\bar{r}),
\end{equation}
so that Eq.~(\ref{h-eq-adim2}) becomes
\begin{equation}
\label{h-eq-adim3}
\frac{\partial \bar{h}}{\partial \bar{t}} = 1 - \frac{\mathcal{P}(\bar{r})}{\overline{V}}+ \frac{\delta^2}{3\overline{V}}\frac{k_{\perp}}{k_{\parallel}} \frac{\bar{L}^3 }{\bar{L}+\bar{h}(\bar{r})}\nabla_{\bar{r}}^2 \mathcal{P}(\bar{r}).
\end{equation}

It is convenient to measure the position of the air-liquid interface with respect to its static initial position. Therefore, we introduce
\begin{equation}
H(\bar{r},\bar{t}) = \bar{h}(\bar{r},\bar{t}) - \bar{h}_J \quad \Rightarrow \quad \bar{h}+\bar{L} = 1 + H + \bar{h}_J -\bar{t}, \quad \mathcal{P} = \frac{H}{1 + H + \bar{h}_J -\bar{t}},
\end{equation}
where we have used $L=L_0-Vt$ and the change of variables (\ref{change-var-sup}). We thus have
\begin{subequations}
\begin{align}
\frac{\partial \bar{h}}{\partial \bar{r}} &= \frac{\partial H}{\partial \bar{r}}, \quad \frac{\partial \mathcal{P}}{\partial \bar{r}} = \frac{(1 +\bar{h}_J -\bar{t})}{(1 + H +\bar{h}_J -\bar{t})^2} \frac{\partial H}{\partial \bar{r}}, \\ 
\nabla_{\bar{r}}^2 \mathcal{P} &=  \frac{\partial^2 \mathcal{P}}{\partial \bar{r}^2} + \frac{1}{\bar{r}} \frac{\partial \mathcal{P}}{\partial \bar{r}} =\frac{(1+\bar{h}_J-\bar{t})}{(1 + H+\bar{h}_J -\bar{t})^2} \nabla_{\bar{r}}^2 H - \frac{2(1+\bar{h}_J-\bar{t})}{(1 + H+\bar{h}_J -\bar{t})^3} \left[\frac{\partial H}{\partial \bar{r}}\right]^2.
\end{align}
\end{subequations}
Substituting these results in Eq.~(\ref{h-eq-adim3}), we finally get
\begin{equation}
\label{h-eq-adim-final}
\frac{\partial H}{\partial \bar{t}} = 1 - \frac{1}{\overline{V}} \left[\frac{H}{1 + H + \bar{h}_J -\bar{t}} \right]+ \frac{\delta^2}{3\overline{V}}\frac{k_{\perp}}{k_{\parallel}} \frac{(1-\bar{t})^3(1+\bar{h}_J- \bar{t})}{(1 + H+\bar{h}_J -\bar{t})^3}\left[ \nabla_{\bar{r}}^2 H - \frac{2}{(1 + H+\bar{h}_J -\bar{t})}(\partial_{\bar{r}} H)^2\right].
\end{equation}
It appears that the last term of Eq.~(\ref{h-eq-adim-final}) is very small and neglecting it leads to Eq.~(9) of the main text: 
\begin{equation}
\label{h-eq-adim-final2}
\frac{\partial H}{\partial \bar{t}} = 1 - \frac{1}{\overline{V}} \left[\frac{H}{1 + H + \bar{h}_J -\bar{t}} \right]+ \frac{\bar{\delta}^2}{3\overline{V}} \frac{(1-\bar{t})^3(1+\bar{h}_J- \bar{t})}{(1 + H+\bar{h}_J -\bar{t})^3} \nabla_{\bar{r}}^2 H,
\end{equation}
where $\bar{\delta}=\delta (k_{\perp}/k_{\parallel})^{1/2}$. Indeed, the relative error on the maximum value of $H(0,t)$ introduced by neglecting this term is at most of 3.5\% when $0 \leq \bar{\delta} \leq 2$, $0 \leq \overline{V} \leq 5$ and $0.1 \leq \bar{h}_J \leq 0.5$. Eq.~(\ref{h-eq-adim-final2}) is solved numerically with the initial condition $H(\bar{r},0)=0$ and the boundary conditions $\partial_{\bar{r}} H(\bar{r},\bar{t})|_{\bar{r} = 0} = H(1,\bar{t}) =0$.

\subsection{Pressure and velocity fields}

\begin{figure*}[!t]
\centering
\includegraphics[width=\textwidth]{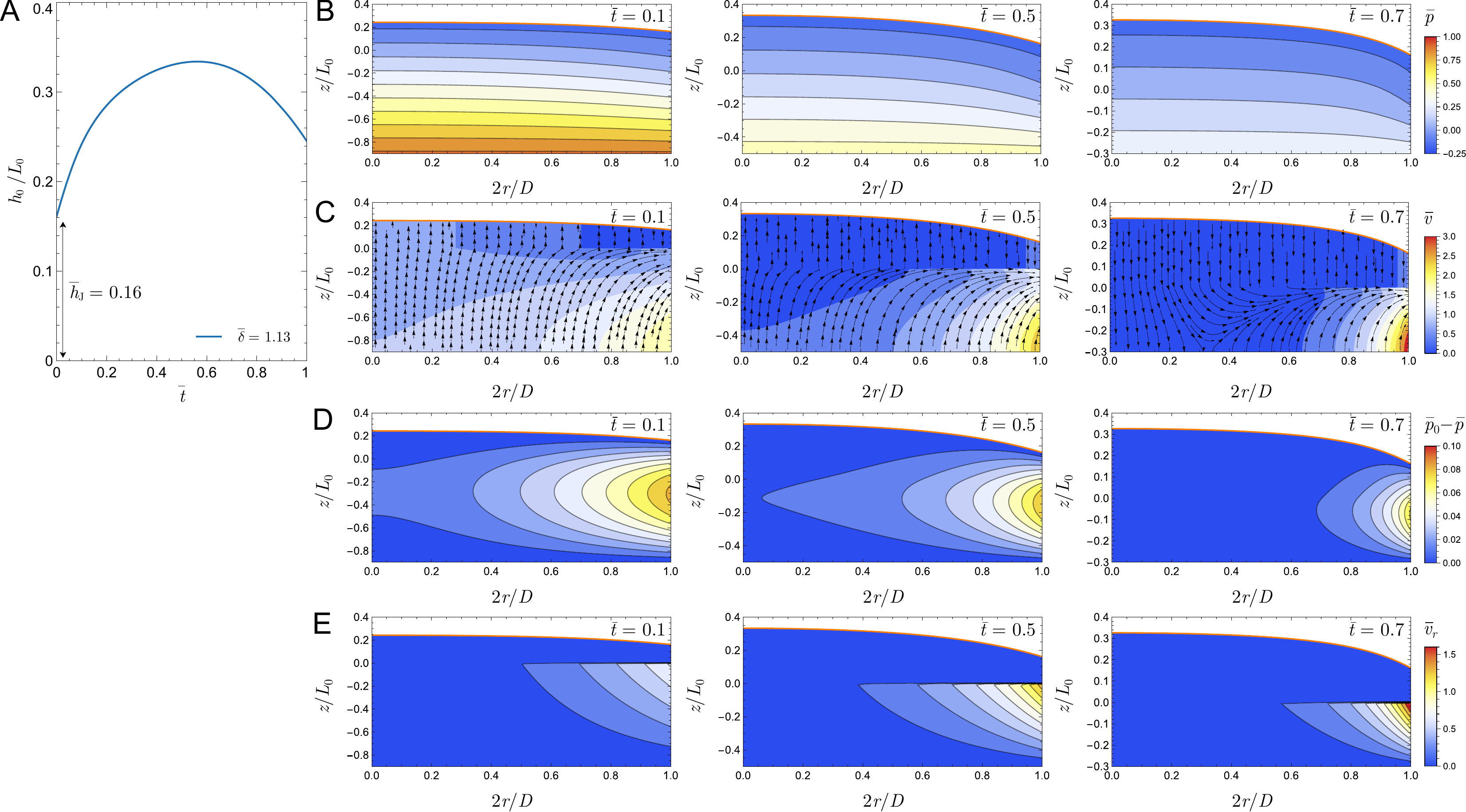}
\caption{(A) Evolution of $h_0/L_0 = H_0 + \bar{h}_J$ as a function of $\bar{t}=Vt/L_0$ for $\overline{V}=0.24$, $\bar{\delta} = 1.13$ and $\bar{h}_J = 0.16$ obtained by solving numerically Eq.~(\ref{h-eq-adim-final2}). Corresponding pressure field $\bar{p}=p/\rho g L_0$ (B) and velocity field $\vec{v}/V$ (C) in the liquid inside the brush at three different times. The density field in panels (C) corresponds to the norm of the velocity $\bar{v}=|\vec{v}|/V$. (D) Pressure field $\bar{p}_0 - \bar{p}$. (E) Transverse component of the velocity field $\bar{v}_r=v_r/V$.} 
\label{FigS-maps-p-v}
\end{figure*}

Fig.~\ref{FigS-maps-p-v}A shows the temporal evolution of the height of the interface $h_0/L_0$ for a typical case where a brush ($R=500$ $\mu$m, $d=2.0$ mm, $D=20$ mm) immersed at a depth $L_0=16$ mm in a silicon oil bath ($\mu=0.97$ Pa s) is removed at a speed $V=16$ mm/min. The corresponding dimensionless parameters used in Eq.~(\ref{h-eq-adim-final2}) are written in Fig.~\ref{FigS-maps-p-v}A. Once $H(\bar{r},\bar{t})$ is obtained, the pressure field is computed using Eq.~(\ref{p0-p1}) [Fig.~\ref{FigS-maps-p-v}B] which is then used in Eq.~(\ref{vr-vz2}) to obtain the velocity field [Fig.~\ref{FigS-maps-p-v}C]. Initially, the isobars are horizontal and the fluid goes in the upward direction until gravity starts draining it out. As soon as $h_0(t)$ increases, the streamlines start bending outward and the flow escapes the brush mostly in the radial direction, just under the surface of the bath, where the pressure gradients are maximal. During this phase, the fluid height reaches its maximum value [Fig.~\ref{FigS-maps-p-v}A]. At later time, the mean flow reverts and the fluid starts going downwards [Fig.~\ref{FigS-maps-p-v}C last panel] thereby causing $h_0(t)$, the height of the interface, to decrease [Fig.~\ref{FigS-maps-p-v}A]. 

From these observations, we see that the evolution of $h_0(t)$ observed in Fig.~\ref{FigS-maps-p-v}A translates the complex time evolution of the underlying flow and pressure fields during the brush withdrawal. Moreover, it is an easy quantity to extract from experimental data and was therefore chosen as the observable we use, in the main text, to compare our model with our experiments.

\section{Additional information about the model}

\subsection{Retraction speed above which $H_0$ grows monotonically in time}

During the retraction of a brush, the maximum height of the liquid interface entrained, $H_0(\bar{t}) = H(0,\bar{t})$, is located in the middle of the brush. As observed in our experiments and in the numerical analysis of Eq.~(\ref{h-eq-adim-final2}), the temporal evolution of $H_0$ can be either non-monotonic, featuring a maximum for $0< \bar{t} < 1$, or monotonic where $H_0$ reaches its highest value for $\bar{t} =1$. 
The transition between these two regimes occurs at some retraction speed, $\overline{V}_{T}$ such that $H_0'(1)=0$, where prime denotes a time derivative. Evaluating Eq.~(\ref{h-eq-adim-final2}) at $\bar{t}=1$ where $\bar{L}=0$ and imposing $H_0'(1)=0$ shows that this transition occurs when $H_0(1) = \overline{V}_{T} \bar{h}_J /(1-\overline{V}_{T})$. Since $H_0(1) \le 1$, (i.e. the interface cannot move faster than the brush), we know that $\overline{V}_{T} \le 1/(1+\bar{h}_J)$. 

\begin{figure*}[!t]
\centering
\includegraphics[width=\textwidth]{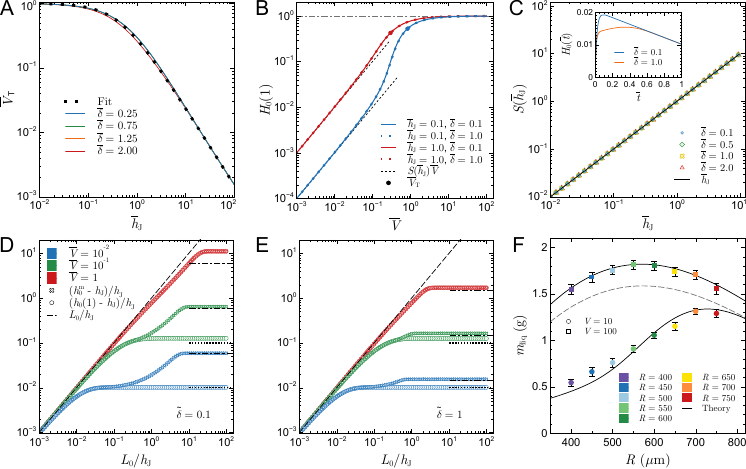}
\caption{(A) Evolution of the transition speed $\overline{V}_{T}$ as a function of $\bar{h}_{J}$ for several values of $\bar{\delta}$. The dotted curve corresponds to the fit (\ref{VT-fit}). (B) Evolution of the height of the interface at the end of the retraction, $H_0(1)$, as a function of $\overline{V}$ for several values of $\bar{h}_J$ and $\bar{\delta}$. The circular dots on the curves indicate the value $\overline{V}_{T}$ of the retraction speed beyond which $H_0^m = H_0(1)$. At low $\overline{V}$, $H_0(1)$ varies linearly with $\overline{V}$, $H_0(1) = S(\bar{h}_J) \overline{V}$. (C) Evolution of $S(\bar{h}_J)$ as a function of $\bar{h}_J$ for several values of $\bar{\delta}$. The numerical data are fitted by $S = \bar{h}_J$. Inset: Temporal evolution of $H_0(\bar{t})$ for $\overline{V}=10^{-2}$, $\bar{h}_J = 1$ and two values of $\bar{\delta}$. (D) Evolution of $(h_0^m-h_J)/h_J$ and $(h_0(1)-h_J)/h_J$ as a function of $L_0/h_J$ for $\tilde{\delta} \equiv \bar{\delta} \bar{h}_J = 0.1$. The dashed curve indicates the short-time behavior whereas the horizontal dashed-dotted and dotted curve are estimations of the saturation values obtained from Eq.~(\ref{Deltah0m-sup}) and Eq.~(\ref{Deltah01}), respectively. (E) Same as panel (D) for $\tilde{\delta} = 1$. (F) Measured mass of liquid, $m_{\text{liq}}$, captured at the end of the retraction process by brushes with various pillar radii $R$ for two retraction speeds (expressed in mm/min) and $d=1.8$ mm, $L_0 = 10$ mm, $\mu = 0.97$ Pa s. The solid curves correspond to the mass computed with the theory. $D=15.8$ mm and $D=16.9$ mm were used in the theory for $V = 10$ mm/min and $V = 100$ mm/min, respectively. The dashed curve corresponds to the theoretical curve at $V = 100$ mm/min when $D=15.8$ mm is used. See text for discussion.} 
\label{FigS-VT}
\end{figure*}

The actual value of $\overline{V}_{T}$, can be computed numerically from Eq.~(\ref{h-eq-adim-final2}). Figure~\ref{FigS-VT}A shows the evolution of $\overline{V}_{T}$ as a function of $\bar{h}_{J}$ for some values of $\bar{\delta}$. It is seen that $\overline{V}_{T}$ is insensitive to the value of $\bar{\delta}$ when it is smaller than about 2. In this case, $\overline{V}_{T}$ can be fitted by
\begin{equation}
\label{VT-fit}
\overline{V}_{T} \approx \frac{1}{1+2.63\, \bar{h}_{J}^{6/5}}, \quad \text{for} \quad 0 \le \bar{h}_{J} \le 10^2.
\end{equation}

\subsection{Variation of $H_0(1)$ with the retraction speed}

In the main text, we have obtained the variation of the maximum height reached by the interface as a function of the retraction speed provided it is smaller than $\overline{V}_{T}$
\begin{equation}
\label{Deltah0m-sup}
H_0^m \simeq [\bar{h}_J+\alpha(\bar{\delta})]\, \overline{V}, \quad \overline{V} \lesssim \overline{V}_{T},
\end{equation}
where $\alpha^2(x)\approx \tanh[1/(2 x)^2]$. A similar result can be obtained for the height of the interface at the end of the retraction, $H_0(1)$. Fig.~\ref{FigS-VT}B shows the evolution of $H_0(1)$ as a function of the retraction speed for several values of $\bar{\delta}$ and $\bar{h}_J$. When $\overline{V} \lesssim \overline{V}_{T}$, $H_0(1)$ evolves linearly with $\overline{V}$. The evolution of the slope with $\bar{h}_J$ is given in Fig.~\ref{FigS-VT}C for several values of $\bar{\delta}$. The overall variation of $H_0(1)$ reads as
\begin{equation}
\label{Deltah01}
H_0(1) \simeq \bar{h}_J\, \overline{V}, \quad \overline{V} \lesssim \overline{V}_{T}.
\end{equation}
In contrast with $H_0^m$, $H_0(1)$ is essentially insensitive to the value of $\bar{\delta}$. This is due to the presence of $\bar{L}^3 = (1-\bar{t})^3$ in the term proportional to $\bar{\delta}$ in Eq.~(\ref{h-eq-adim-final2}) which vanishes quickly as $\bar{t}\to 1$. The inset of Fig.~\ref{FigS-VT}C shows the effect of a variation of $\bar{\delta}$ by one order of magnitude on the temporal evolution of $H_0(\bar{t})$. It is seen that $H_0^m$ changes significantly but $H_0(1)$ remains essentially unchanged. 

\subsection{Influence of the immersion depth}

The influence of the retraction speed on $H_0^m$ and $H_0(1)$ has been well characterized. At small retraction speed, their expression are given by Eq.~(\ref{Deltah0m-sup}) and Eq.~(\ref{Deltah01}). At large speed, they both tend to 1. We discuss briefly here the influence of the immersion depth on this two quantities.

Figs.~\ref{FigS-VT}D and E show the evolution of $(h_0^m-h_J)/h_J$ and $(h_0(1)-h_J)/h_J$ as a function of $L_0/h_J$ for two values of $\tilde{\delta} \equiv \bar{\delta} \bar{h}_J$. We use here $h_J$ to rescaled $h_0-h_J$ instead of $L_0$ because $H_0^m$ and $H_0(1)$ would vary with $L_0$ for fixed values of $h_0^m$ and $h_0(1)$. We also use $\tilde{\delta}$ instead of $\bar{\delta}$ because the former does not depend on $L_0$. In other words, we use $h_J$ as a vertical length scale instead of $L_0$ in the change of variables Eq.~(\ref{change-var-sup}). 

These two figures show that, at small $L_0/h_J\equiv \bar{h}_J^{-1}$, we have $h_0^m=h_0(1)$. Indeed, as shown above, $h_0^m=h_0(1)$ only when $\overline{V} >\overline{V}_{T}$. For a given value of $\overline{V} < 1$, this happens only for $\bar{h}_J$ large enough (see Fig.~\ref{FigS-VT}A and Eq.~(\ref{VT-fit})) or, equivalently, for $L_0/h_J$ sufficiently small. It is also seen that $h_0^m=h_0(1)\simeq h_J + L_0$ when $L_0/h_J \ll 1$. Indeed, as just discussed, $L_0/h_J \ll 1$ is equivalent to $\overline{V} \gg \overline{V}_{T}$ and in this case, the liquid moves at the same speed than the pillars, i.e. $h_0 = h_J + Vt$, and at the end of retraction ($t = L_0/V$), we have $h_0 = h_J + L_0$. In other words, for small immersion depth, the time needed to completely remove a brush from a bath is small and the liquid has no time to flow out of the brush and just follow the pillars.

Figs.~\ref{FigS-VT}D and E also show, as expected, that the height of the interface saturates to a maximum value when $L_0$ is sufficiently large ($L_0 \gtrsim h_J$). The saturation value for $h_0(1)$ can be estimated from Eq.~(\ref{Deltah01}) which can be written as: $(h_0(1)-h_J)/h_J= \overline{V}$. Each dotted line in Figs.~\ref{FigS-VT}D and E corresponds thus to $\overline{V}$. Similarly, the saturation value for $h_0^m$ can be estimated from Eq.~(\ref{Deltah0m-sup}) which can be written as: $(h_0^m-h_J)/h_J= [1+\alpha(\tilde{\delta}/\bar{h}_J)/\bar{h}_J] \overline{V}$. In the limit $L_0/h_J \gg 1$, i.e. $\bar{h}_J \ll 1$, at fixed $\tilde{\delta}$, the function $\alpha$ can be expanded using its approximate expression
\begin{equation}
\label{alpha-eq}
\alpha(x)\approx \left[\tanh[1/(2 x)^2]\right]^{1/2} \simeq \frac{1}{2x} \quad \text{for}\quad x \gg 1.
\end{equation}
Using this result, we have
\begin{equation}
\frac{h_0^m-h_J}{h_J}= \left[1+\frac{\alpha(\tilde{\delta}/\bar{h}_J)}{\bar{h}_J}\right] \overline{V} \simeq \left[1+\frac{1}{2\tilde{\delta}}\right] \overline{V}.
\end{equation}
This last expression is shown as horizontal dashed-dotted lines in Figs.~\ref{FigS-VT}D and E.

\subsection{Discussion about the value of $D$ used to compute the mass of liquid captured} 

The value of the brush diameter $D$ affects significantly the (dimensional) mass of liquid computed from Eq.~(12) of the main text since $\mathcal{V}$ is proportional to $D^2$ whereas the impact of $D$ on $H$ computed from Eq.~(\ref{h-eq-adim-final2}) is much less pronounced, especially at the end of retraction. Indeed, $D$ appears only in the expression of $\bar{\delta}$ in Eq.~(\ref{h-eq-adim-final2}) which has a negligible influence on profile of $H$ at the end of the retraction as seen in Fig.~\ref{FigS-VT}B where $\bar{\delta}$ varies by one order of magnitude. Therefore, $D^2$ acts as a scaling factor in the expression of $m_{\text{liq}} = \rho \mathcal{V}$. However, $D$ is not well defined for our brushes, see Fig.~1B of the main text, but it is expected to grow with the retraction speed as more liquid is then captured at the rim of the brush. The brush diameter could be defined as the diameter of the circle passing through the centers of the pillars at the rim of the brush (as shown in Fig.~1B of the main text and $D= 15.8$ mm for the brushes used in Fig.~4C of the main text) or passing through the borders of the pillars at the rim of the brush ($D' = D + 2R$, $16.6 \le D' \le 17.4$ mm for these brushes). We found a good agreement with the data in Fig.~4C of the main text using $D=15.8$ mm and $D = 16.9$ mm for $V = 10$ mm/min and $V = 100$ mm/min, respectively. The dashed curve in Fig.~\ref{FigS-VT}F shows how the theoretical curve is affected when $D=15.8$ mm is used for $V = 100$ mm/min. The value of the optimal radius predicted by the theory is, of course, unaffected by a change of a multiplicative constant, i.e. $D^2$, in the expression of $m_{\text{liq}}$.

\section{Approximate solutions of the PDE (\ref{h-eq-adim-final2})}

\subsection{Values of $\bar{\delta}$ below which the solutions with $\bar{\delta}=0$ leads to good approximations}

The PDE (\ref{h-eq-adim-final2}) reduces to an ODE when $\bar{\delta}=0$ which is simpler to deal with: 
\begin{equation}
\label{h-eq-ODE}
\frac{d H_0^{1D}}{d \bar{t}} = 1 - \frac{1}{\overline{V}} \left[\frac{H_0^{1D}}{1 + H_0^{1D} + \bar{h}_J -\bar{t}} \right], \quad H_0^{1D}(0) = 0.
\end{equation}
Here, we analyze under which conditions the temporal evolution of $H_0(\bar{t}) = H(0,\bar{t})$ is well approximated by $H_0^{1D}(\bar{t})$. For this purpose, we define
\begin{equation}
\label{eps-def}
\varepsilon(\overline{V}, \bar{h}_J, \bar{\delta}) = \frac{H_0^{1D,m}-H_0^m}{H_0^m},
\end{equation}
where $H_0^{1D,m}$ and $H_0^m$ are the largest value of $H_0^{1D}$ and $H_0$ in the interval $0 \le \bar{t} \le 1$ respectively. $\varepsilon$ is thus the relative error on the largest value of $H_0$ and is always positive because $H_0^{1D,m}\ge H_0^m$. Now, we define $\bar{\delta}_0$ such as 
\begin{equation}
\label{del0-def}
\varepsilon(\overline{V}, \bar{h}_J, \bar{\delta}_0) = 0.01, \quad \Rightarrow \quad \bar{\delta}_0 = \bar{\delta}_0(\overline{V}, \bar{h}_J).
\end{equation}
Therefore, for a given system where $\overline{V}$, $\bar{h}_J$ and $\bar{\delta}$ are known, if $\bar{\delta} \le \bar{\delta}_0(\overline{V}, \bar{h}_J)$, then the relative error on $H_0^m$ obtained by using Eq.~(\ref{h-eq-ODE}) instead of Eq.~(\ref{h-eq-adim-final2}) is smaller or equal to 1\%.

\begin{figure*}[!t]
\centering
\includegraphics[width=\textwidth]{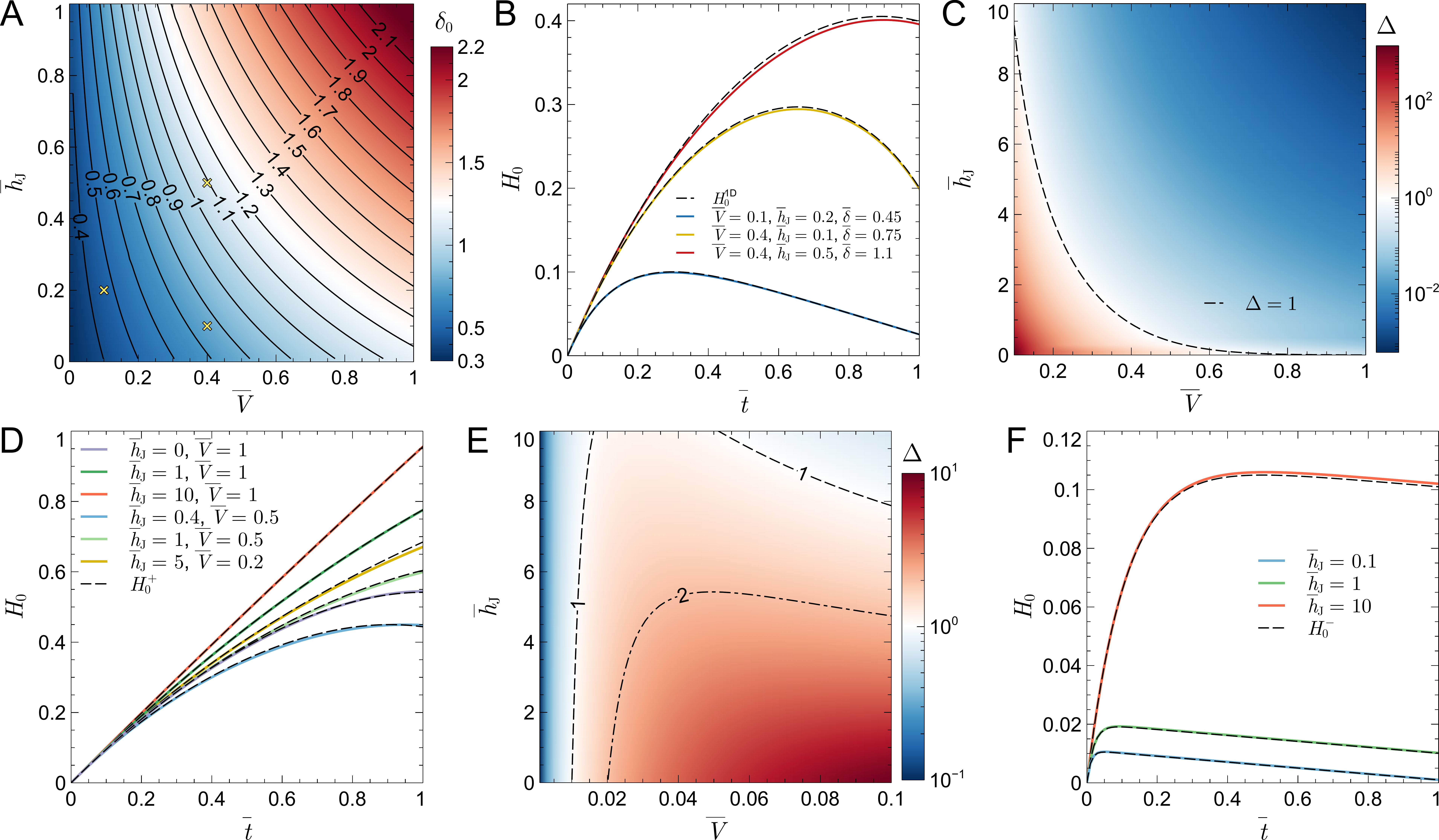}
\caption{(A) Evolution of $\bar{\delta}_0$, defined in Eqs.~(\ref{eps-def}) and (\ref{del0-def}), as a function of $\overline{V}$ and $\bar{h}_J$. The yellow crosses shows illustrative cases shown in panel (B). (B) Temporal evolution of $H_0$ computed with Eq.~(\ref{h-eq-adim-final2}) for various values of $\overline{V}$ and $\bar{h}_J$ and $\bar{\delta} \simeq \bar{\delta}_0$. The dashed lines show the temporal evolution of $H_0^{1D}$ computed with Eq.~(\ref{h-eq-ODE}). (C) Evolution of $\Delta$, defined in Eq.~(\ref{Delta-def}), as a function of $\overline{V}$ and $\bar{h}_J$ for $\bar{\delta}=1$. (D) Comparison between the temporal evolution of $H_0$ computed with Eq.~(\ref{h-eq-adim-final2}) and the approximate expression $H_0^+$, defined in Eq.~(\ref{approx-large-V}), for various values of $\overline{V}$ and $\bar{h}_J$ and $\bar{\delta}=1$. In these examples, $\Delta < 0.27$ for $\overline{V} = 1$, $\Delta < 0.73$ for $\overline{V} = 0.5$ and $\Delta = 0.53$ for $\overline{V} = 0.2$. (E) Evolution of $\Delta$, defined in Eq.~(\ref{Delta-def}) with $H_0^{+}$ replaced by $H_0^{-}$, as a function of $\overline{V}$ and $\bar{h}_J$ for $\bar{\delta}=1$. $H_0^{-}$ is given by Eq.~(\ref{approx-small-V}). (F) Comparison between the temporal evolution of $H_0$ computed with Eq.~(\ref{h-eq-adim-final2}) and the approximate expression $H_0^-$ for various values of $\bar{h}_J$, $\overline{V}=0.01$ and $\bar{\delta}=0.1$. In these examples, $0.76 \le \Delta \le 0.99$.}
\label{FigS-approx-sol}
\end{figure*}

Fig.~\ref{FigS-approx-sol}A shows the evolution of $\bar{\delta}_0$ as a function of $\overline{V}$ and $\bar{h}_J$. The three cross indicates three illustrative cases shown in Fig.~\ref{FigS-approx-sol}B. For example, when $\overline{V}=0.1$ and $\bar{h}_J=0.1$, $\bar{\delta}_0 \simeq 0.45$. Therefore, if $\bar{\delta} \le 0.45$, the evolution of $H_0$ is well captured by $H_0^{1D}$ obtained from the ODE (\ref{h-eq-ODE}) as seen in Fig.~\ref{FigS-approx-sol}B.

\subsection{Analytical approximate solution for large retraction speed} In this regime, $1/\overline{V} \ll 1$ and we expand $H$ as follows:
\begin{equation}
H(\bar{r},\bar{t}) = \sum_{i=0} \overline{V}^{-i} \, H^{(i)}(\bar{r},\bar{t}).
\end{equation}
Substituting this expansion into Eq.~(\ref{h-eq-adim-final2}) and solving order by order, we get at order $\overline{V}^0$, $H^{(0)}=\bar{t}$ as expected since this is the evolution of $H^{(0)}$ for an infinite retraction speed. Because $H^{(0)}$ does not depend on $\bar{r}$, the others $H^{(i>0)}$ are only function of $\bar{t}$. Therefore, in this approximation scheme, we have $H(\bar{r},\bar{t}) = H_0(\bar{t})$. Up to order $\overline{V}^{-2}$, we find
\begin{equation}
\label{approx-large-V}
H_0(\bar{t}) \simeq H_0^+(\bar{t}) = t - \frac{1}{2 \overline{V} (1 + \bar{h}_J)}\, t^2 + \frac{4 + 4 \bar{h}_J - 3 t}{24\overline{V}^2 (1 + \bar{h}_J)^3} t^3.
\end{equation}
To estimate the accuracy of this approximate solution, we define
\begin{equation}
\label{Delta-def}
\Delta = 100 \int_0^1 \left[\frac{|H_0(\bar{t}) - H_0^+(\bar{t})|}{H_0(\bar{t})} \, d\bar{t}\right],
\end{equation}
so that $H_0^+$ gives a good approximation of $H_0$ when $\Delta \lesssim 1$. Fig.~\ref{FigS-approx-sol}C shows the evolution of $\Delta$ as a function of $\overline{V}$ and $\bar{h}_J$ for $\bar{\delta}=1$. Fig.~\ref{FigS-approx-sol}D shows some comparisons between $H_0(\bar{t})$ computed with Eq.~(\ref{h-eq-adim-final2}) and $H_0^+(\bar{t})$ defined in Eq.~(\ref{approx-large-V}). In these examples, $H_0^+$ approximates well $H_0$ because $7\times 10^{-4} \le \Delta \le 0.73$. 

\subsection{Analytical approximate solution for small retraction speed} As shown above, when $\bar{\delta} < \bar{\delta}_{0}(\overline{V}, \bar{h}_J)$, see Eq.~(\ref{del0-def}), $H_0^{1D}$ obtained by solving Eq.~(\ref{h-eq-ODE}) is a good approximation of the full solution $H_0$. Here, we further assume that $\overline{V}\ll 1$ so that $H_0 \ll 1$. In this case, linearizing Eq.~(\ref{h-eq-ODE}) leads to
\begin{equation}
\label{h-eq-ODE-lin}
\frac{d H_0^{-}}{d \bar{t}} = 1 - \frac{1}{\overline{V}} \left[\frac{H_0^{-}}{1 +  \bar{h}_J -\bar{t}} \right], \quad H_0^{-}(0) = 0,
\end{equation}
which can be solved exactly:
\begin{equation}
\label{approx-small-V}
H_0^{-}(\bar{t}) = \frac{\overline{V}}{1-\overline{V}} [1 +  \bar{h}_J -\bar{t}]\left[1-\left(\frac{1 +  \bar{h}_J -\bar{t}}{1+\bar{h}_J}\right)^{(1-\overline{V})/\overline{V}}\right].
\end{equation}
Fig.~\ref{FigS-approx-sol}E shows the evolution of $\Delta$, defined in Eq.~(\ref{Delta-def}) with $H_0^{+}$ replaced by $H_0^{-}$, as a function of $\overline{V}$ and $\bar{h}_J$ for $\bar{\delta}=0.1$. Fig.~\ref{FigS-approx-sol}F shows some comparisons between $H_0(\bar{t})$ computed with Eq.~(\ref{h-eq-adim-final2}) and $H_0^-(\bar{t})$ defined in Eq.~(\ref{approx-small-V}). In these examples, $H_0^-$ approximates well $H_0$ because $0.76 \le \Delta \le 0.99$.

\section{Parallel plates}

An equation similar to Eq.~(\ref{h-eq-adim-final2}) can be obtained in Cartesian coordinates for two parallel plates separated by a distance $2d$ along the $x$-axis, as shown in Fig.~\ref{FigS-Jurin}A. For this purpose, we first compute the permeability before to derive the PDE.

\subsection{Permeability}

We consider here the flow between two immobile parallel plates as shown in Fig.~\ref{FigS-Jurin}A. Since the $x$-direction is the confinement direction the lubrication equations for a Newtonian and incompressible fluid read in Cartesian coordinates as
\begin{equation}
\label{eqs-vz-vy}
\mu \frac{\partial^2 u_z}{\partial x^2} = \frac{\partial p}{\partial z} + \rho g, \quad \mu \frac{\partial^2 u_y}{\partial x^2} = \frac{\partial p}{\partial y},\quad \frac{\partial p}{\partial x} = 0,\quad  \frac{\partial u_x}{\partial x} + \frac{\partial u_y}{\partial y} + \frac{\partial u_z}{\partial z}=0. 
\end{equation}
Symmetry at $x=0$ and non-slip at the walls ($x=d$) yields the following boundary conditions
\begin{equation}
\label{BC-plates}
u_x = \frac{\partial u_y}{\partial x} = \frac{\partial u_z}{\partial x} = 0 \quad \text{at}\quad x = 0, \quad \quad u_z = u_y = u_x =  0 \quad \text{at}\quad x = d.
\end{equation}
The third of Eqs.~(\ref{eqs-vz-vy}) yields $p = p(y,z)$. The first and second of Eqs.~(\ref{eqs-vz-vy}) can then be easily integrated using the boundary conditions (\ref{BC-plates}) for $u_z$ and $u_y$:
\begin{equation}
\label{vy-vz-sol}
u_z(x,y,z) = \frac{(x^2-d^2)}{2\mu} \left(\frac{\partial p}{\partial z} + \rho g\right), \quad u_y(x,y,z) = \frac{(x^2-d^2)}{2\mu} \frac{\partial p}{\partial y}.
\end{equation}
Mass conservation, together with the boundary condition (\ref{BC-plates}) for $u_x$ at $x=0$ leads to
\begin{equation}
u_x = \frac{(3d^2 x - x^3)}{6\mu} \left(\frac{\partial^2 p}{\partial y^2}+ \frac{\partial^2 p}{\partial z^2}\right).
\end{equation}
Finally, the boundary condition (\ref{BC-plates}) at $x=d$ for $u_x$ yields the following equation for the pressure
\begin{equation}
\frac{\partial^2 p}{\partial y^2}+ \frac{\partial^2 p}{\partial z^2}=0.
\end{equation}
This equation must be solved for $x=d$ but, because $p$ does not depend on $x$, it is valid for any $x$ and $u_x\equiv 0$. Because $p$ does not depend on $x$ and $u_x = 0$, it is natural to consider the $x$-averaged velocity field. We define
\begin{equation}
v_i(y,z) = \frac{1}{2d}\int_{-d}^{d} u_i(x,y,z)\, dx.
\end{equation}
We then obtain an effective two-dimensional problem with the following velocity field
\begin{equation}
\label{vy-vz-sol-av}
v_y(y,z) = -\frac{d^2}{3\mu} \frac{\partial p}{\partial y} \equiv -\frac{k_{\perp}}{\mu} \frac{\partial p}{\partial y}, \quad
v_z(y,z) = -\frac{d^2}{3\mu} \left(\frac{\partial p}{\partial z} + \rho g\right) \equiv - \frac{k_{\parallel}}{\mu} \left(\frac{\partial p}{\partial z} + \rho g\right).
\end{equation}
The velocity is thus given by Darcy's law with the permeabilities $k_{\perp}/d^2 = k_{\parallel}/d^2 = 1/3$.

\subsection{Pressure and velocity fields between two parallel plates removed at constant speed from a bath}

The velocity and pressure fields of the fluid between two parallel plates separated by a distance $2d$ along the $x$-axis and removed along the $z$-axis at speed $V$ from a liquid bath is given by Darcy's law in Cartesian coordinates 
\begin{equation}
\label{vy-vz-dim}
v_{y}(y,z) = -\frac{k_{\perp}(z)}{\mu} \frac{\partial p}{\partial y}, \quad
v_z(y,z) = V - \frac{k_{\parallel}}{\mu} \left(\frac{\partial p}{\partial z} + \rho g\right), \quad \frac{\partial^2 p}{\partial z^2} +\frac{k_{\perp}(z)}{k_{\parallel}}\frac{\partial^2 p}{\partial y^2} =0,
\end{equation}
where the equation for the pressure is obtained from mass conservation and $z=0$ corresponds to the position of the air-liquid interface of the bath. As for brushes, we consider that there is no flow along the transverse (horizontal) direction inside the fluid transported by the brush located above the level of the liquid bath. Therefore
\begin{equation}
\label{kpz-plate}
k_{\perp}(z) = k_{\perp} \theta(-z) = \frac{d^2}{3} \theta(-z), \qquad k_{\parallel}=\frac{d^2}{3}.
\end{equation}
where $\theta$ is the Heaviside function. Using the change of variables
\begin{equation}
\label{change-var-plate}
\bar{y} = 2y/W, \quad (\bar{z},\bar{h},\bar{L},\bar{h}_{J},\bar{t},\bar{p}) = (z,h,L,h_J,V t, p/\rho g)/L_0,
\end{equation}
where $W$ is the width of the plates along the $y$-axis, the equation for the pressure becomes
\begin{equation}
\label{eq-p-adim-plate}
\frac{\partial^2 \bar{p}}{\partial \bar{z}^2}+ \delta^2 \frac{k_{\perp}(\bar{z})}{k_{\parallel}} \frac{\partial^2 \bar{p}}{\partial \bar{y}^2} =0, \quad \delta = \frac{2L_0}{W},
\end{equation}
where the aspect ratio $\delta$ is assumed to be small. Expanding the pressure up to order $\delta^2$, $\bar{p} = \bar{p}_0 + \delta^2\, \bar{p}_1$, 
Eq.~(\ref{eq-p-adim-plate}) becomes at orders $\delta^0$ and $\delta^2$
\begin{equation}
\label{eq-p0-p1-plate}
\frac{\partial^2 \bar{p}_0}{\partial \bar{z}^2}=0, \quad \frac{\partial^2 \bar{p}_1}{\partial \bar{z}^2}=-\frac{k_{\perp}(\bar{z})}{k_{\parallel}} \frac{\partial^2 \bar{p}_0}{\partial \bar{y}^2}.
\end{equation}
These equations are solved with the boundary conditions
\begin{equation}
\bar{p}_0(\bar{y},-\bar{L}) = \bar{L}, \quad \bar{p}_0(\bar{y},\bar{h}) = -\bar{h}_{J}, \quad \bar{p}_1(\bar{y},-\bar{L}) = 0, \quad \bar{p}_1(\bar{y},\bar{h}) = 0.
\end{equation}
The solutions are
\begin{subequations}
\label{p0-p1-plate}
\begin{align}
\bar{p}_0(\bar{y},\bar{z}) &= (\mathcal{P}(\bar{y})-1)\, \bar{z} + \bar{L} \mathcal{P}(\bar{y}), \quad \mathcal{P} = \frac{\bar{h}(\bar{y})-\bar{h}_{J}}{\bar{h}(\bar{y})+\bar{L}}, \\
\bar{p}_1(\bar{y},\bar{z}) &= \frac{1}{6} \left[-\bar{z}^2\left(\bar{z} + 3 \bar{L}\right)\theta(-\bar{z}) + \frac{2\bar{L}^3(\bar{h}(\bar{y})-\bar{z})}{\bar{h}(\bar{y}) + \bar{L}}\right] \frac{\partial^2\mathcal{P}(\bar{y})}{\partial \bar{y}^2}.
\end{align}
\end{subequations}
Using the changes of variables (\ref{change-var-plate}), Eq.~(\ref{vy-vz-dim}) for the velocities and the spatio-temporal evolution of the interface are given by
\begin{subequations}
\begin{align}
\label{vy-vz-plate}
v_{y}(\bar{y},\bar{z}) &= -\delta\, \frac{k_{\perp}(\bar{z})}{k_{\parallel}} \, V_{\parallel} \frac{\partial \bar{p}}{\partial \bar{y}}, \quad
v_z(\bar{y},\bar{z}) = V - V_{\parallel} \left(\frac{\partial \bar{p}}{\partial \bar{z}} + 1\right), \quad V_{\parallel}=\frac{k_{\parallel} \rho g}{\mu}, \\
\label{h-eq-adim-plate}
\frac{\partial \bar{h}(\bar{y},\bar{t})}{\partial \bar{t}} &= \frac{v_z(\bar{y},\bar{h})}{V} - \delta \frac{v_y(\bar{y},\bar{h})}{V} \frac{\partial \bar{h}(\bar{y},\bar{t})}{\partial \bar{y}} = \frac{v_z(\bar{y},\bar{h})}{V},
\end{align}
\end{subequations}
where we used $v_y(\bar{y},\bar{h})=0$ because $k_{\perp}(\bar{h})=0$ since $\bar{h}>0$, see Eqs.~(\ref{vy-vz-dim}) and (\ref{kpz-plate}).
Substituting Eq.~(\ref{vy-vz-plate}) into Eq.~(\ref{h-eq-adim-plate}), we get
\begin{equation}
\label{h-eq-adim2-plate}
\frac{\partial \bar{h}}{\partial \bar{t}} = 1 - \frac{1}{\overline{V}} \left(\frac{\partial \bar{p}}{\partial \bar{z}} + 1\right)_{\bar{z} = \bar{h}} = 1 - \frac{1}{\overline{V}} \left(\frac{\partial \bar{p_0}}{\partial \bar{z}} + \delta^2 \frac{\partial \bar{p_1}}{\partial \bar{z}} + 1\right)_{\bar{z} = \bar{h}}, \quad \text{with} \quad \overline{V} = \frac{V}{V_{\parallel}}.
\end{equation}
where we used the expansion $\bar{p} = \bar{p}_0 + \delta^2\, \bar{p}_1$. This equation is identical to Eq.~(\ref{h-eq-adim2}), only the expression of the pressure is slightly different. Using Eq.~(\ref{p0-p1-plate}), we have
\begin{equation}
\left. \frac{\partial \bar{p}_0}{\partial \bar{z}}\right|_{\bar{z} = \bar{h}} = \mathcal{P}-1, \quad \left. \frac{\partial \bar{p}_1}{\partial \bar{z}}\right|_{\bar{z} = \bar{h}} = - \frac{1}{3}\frac{\bar{L}^3}{\bar{L}+ \bar{h}(\bar{y})} \, \frac{\partial^2\mathcal{P}(\bar{y})}{\partial \bar{y}^2},
\end{equation}
so that Eq.~(\ref{h-eq-adim2-plate}) becomes
\begin{equation}
\label{h-eq-adim4-plate}
\frac{\partial \bar{h}}{\partial \bar{t}} = 1 - \frac{\mathcal{P}}{\overline{V}}+ \frac{\delta^2}{3\overline{V}} \frac{\bar{L}^3}{\bar{L}+ \bar{h}(\bar{y})} \frac{\partial^2\mathcal{P}(\bar{y})}{\partial \bar{y}^2}.
\end{equation}

To measure the position of the air-liquid interface with respect to its static initial position, we introduce
\begin{equation}
H(\bar{y},\bar{t}) = \bar{h}(\bar{y},\bar{t}) - \bar{h}_J \quad \Rightarrow \quad \bar{h}+\bar{L} = 1 + H + \bar{h}_J -\bar{t}, \quad \mathcal{P} = \frac{H}{1 + H + \bar{h}_J -\bar{t}},
\end{equation}
we thus have
\begin{equation}
\frac{\partial \bar{h}}{\partial \bar{y}} = \frac{\partial H}{\partial \bar{y}}, \quad \frac{\partial \mathcal{P}}{\partial \bar{y}} = \frac{(1 + \bar{h}_J-\bar{t})}{(1 + H+ \bar{h}_J -\bar{t})^2} \frac{\partial H}{\partial \bar{y}}, \qquad  
\frac{\partial^2\mathcal{P}(\bar{y})}{\partial \bar{y}^2}=\frac{(1 + \bar{h}_J-\bar{t})}{(1 + H + \bar{h}_J -\bar{t})^2} \left[\frac{\partial^2 H}{\partial \bar{y}^2} - \frac{2}{1 + H+ \bar{h}_J -\bar{t}} \left[\frac{\partial H}{\partial \bar{y}}\right]^2 \right].
\end{equation}
Substituting these results in Eq.~(\ref{h-eq-adim4-plate}), we finally get
\begin{equation}
\label{h-eq-adim-final-plate}
\frac{\partial H}{\partial \bar{t}} = 1 - \frac{1}{\overline{V}} \left[\frac{H}{1 + H+ \bar{h}_J -\bar{t}} \right]+ \frac{\delta^2}{3\overline{V}}\frac{(1-\bar{t})^3 (1 + \bar{h}_J-\bar{t})}{(1 + H + \bar{h}_J -\bar{t})^3} \left[\frac{\partial^2 H}{\partial \bar{y}^2} - \frac{2}{1 + H+ \bar{h}_J -\bar{t}} \left[\frac{\partial H}{\partial \bar{y}}\right]^2 \right].
\end{equation}
As for brushes, the last term of Eq.~(\ref{h-eq-adim-final-plate}) is very small and can be neglected to obtain
\begin{equation}
\label{h-eq-adim-final-plate2}
\frac{\partial H}{\partial \bar{t}} = 1 - \frac{1}{\overline{V}} \left[\frac{H}{1 + H+ \bar{h}_J -\bar{t}} \right]+ \frac{\delta^2}{3\overline{V}}\frac{(1-\bar{t})^3 (1 + \bar{h}_J-\bar{t})}{(1 + H + \bar{h}_J -\bar{t})^3} \frac{\partial^2 H}{\partial \bar{y}^2}.
\end{equation}
Eq.~(\ref{h-eq-adim-final-plate2}) is solved numerically with the initial condition $H(\bar{y},0)=0$ and the boundary conditions $\partial_{\bar{y}} H(\bar{y},\bar{t})|_{\bar{y} = 0} = H(1,\bar{t}) =0$. Once $H$ is known, it is substituted in Eq.~(\ref{p0-p1-plate}) to obtain the pressure field and in Eq.~(\ref{vy-vz-plate}) to obtain the velocity field.

\begin{figure*}[!t]
\centering
\includegraphics[width=0.9\textwidth]{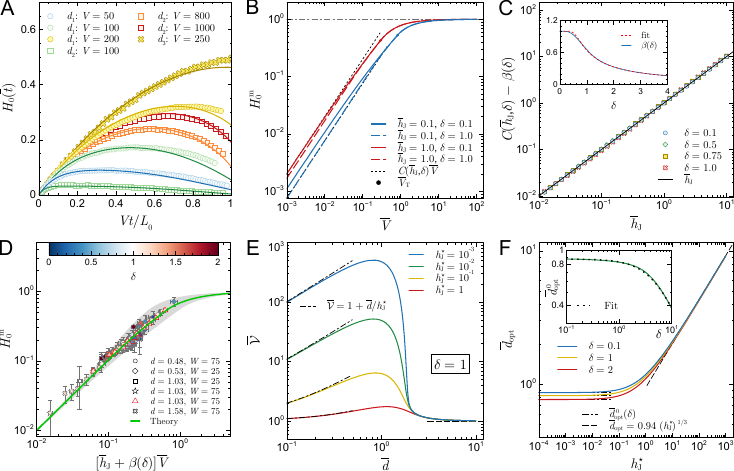}
\caption{(A) Comparison between some temporal variations of $H_0(\bar{t})$ measured experimentally (symbols) and computed from Eq.~(\ref{h-eq-adim-final-plate2}) (solid curves) for two parallel plates: $d_1 = 0.48$ mm ($W= 75$ mm, $L_0 = 40$ mm), $d_2 = 1.03$ mm ($W= 75$ mm, $L_0 = 42$ mm), $d_3 = 0.53$ mm ($W= 25$ mm, $L_0 = 10$ mm), $0.8 \le  \delta \le 1.12$ and $\mu = 0.097$ Pa s. The retraction speed $V$ is given in mm/min.  (B) Evolution of the largest value of $H_0$ in the time interval $0 \le \bar{t} \le 1$, $H_0^m$, as a function of $\overline{V}$ for several values of $\bar{\delta}$ and $\bar{h}_J$ and computed from Eq.~(\ref{h-eq-adim-final-plate2}). The circular dots on the curves indicate the value $\overline{V}_{T}$ of the retraction speed beyond which $H_0^m$ is reached at $\bar{t}=1$. The expression of $\overline{V}_{T}$ for plates is essentially the same than the one for brushes for $\delta \lesssim 1.5$. At low $\overline{V}$, $H_0^m$ varies linearly with $\overline{V}$, $H_0^m = C(\bar{h}_J,\delta) \overline{V}$. (C) Evolution of $C(\bar{h}_J,\delta)$ as a function of $\bar{h}_J$ for several values of $\delta$. The numerical data are fitted by $C = \bar{h}_J + \beta(\delta)$. Inset: Evolution of $\beta$ as a function of $\delta$ together with the fit $\beta(x) \approx \alpha(3 x/4)$ where $\alpha$ is the function appearing when brushes are considered, see Eq.~(\ref{alpha-eq}) and Fig.~2F of the main text. (D) Evolution of $H_0^m$ measured experimentally (symbols) as a function of a rescaled retraction speed $(\bar{h}_J + \beta(\delta))\overline{V}$ for two parallel plates where $20 \le V \le 2000$ mm/min, $5\le L_0 \le 65$ mm and $\mu=0.097$ Pa s except for the data with red edges where $\mu=0.97$ Pa s. The dimensionless control parameters vary in the range: $0.024 \le \overline{V} \le 0.96$, $0.13 \le \delta \le 2.0$, $0.023 \le \bar{h}_J \le 0.92$. The green curve is computed using the average value of $\delta$ and $\bar{h}_J$: $\delta=1.0$ and $\bar{h}_J=0.15$. (E) Evolution of the rescaled volume captured by two parallel plates, $\overline{\mathcal{V}}$, defined in Eq.~(\ref{Volume-plates}), as a function of their rescaled distance $\bar{d}$ for $\delta=2L_0/W =1$ and several values of $h^{\star}_J$. Both $\bar{d}$ and $h^{\star}_J$ are defined in Eq.~(\ref{new-parameters-plates}). (F) Evolution of the optimal distance, $\bar{d}_{\text{opt}}$, between two plates corresponding to the maximum of $\overline{\mathcal{V}}$ shown in panel (D), as a function $h^{\star}_J$ for several values of $\delta$. Inset: Evolution of $\bar{d}_{\text{opt}}^{\,0} = \bar{d}_{\text{opt}}(h^{\star}_J \to 0)$ as a function of $\delta$ together with the fit $\bar{d}_{\text{opt}}^{\,0} \approx \sqrt{3}/2(1+0.075\, \delta^{3/2})^{2/3}$. } 
\label{FigS-plates}
\end{figure*}

\subsection{Comparison with experiments} 

Figs.~\ref{FigS-plates}A shows a comparison between some typical temporal evolution of $H_0$ obtained experimentally with parallel plates, and the corresponding theoretical evolution obtained by solving numerically Eq.~(\ref{h-eq-adim-final-plate2}). A good agreement between theory and experiments is observed for various retraction speeds and immersion depths even when $\delta$ is of order 1. Fig.~\ref{FigS-plates}A shows that the experimental data are quite intricate since the largest values of $H_0^m$ are not necessarily reached for the largest retraction speeds or immersion depths according to the distance between the plates. This complexity is related to the fact that the dynamics is governed by 3 dimensionless groups of parameters: $\overline{V}$, $\bar{h}_J$ and $\delta$.

To get insight into the variation of $h_0^m-h_J$ as a function of the retraction speed, we analyze the behavior of this quantity using Eq.~(\ref{h-eq-adim-final-plate2}). Fig.~\ref{FigS-plates}B shows the evolution of $H_0^m$ as a function of the retraction speed for several values of $\delta$ and $\bar{h}_J$. When $\overline{V} \lesssim \overline{V}_{T}$, i.e. when this maximal value is reached at $\bar{t} < 1$, $H_0^m$ evolves linearly with $\overline{V}$. The evolution of the slope with $\bar{h}_J$ is given in Fig.~\ref{FigS-plates}C for several values of $\delta$. The overall variation of $H_0^m$ involves thus the 3 dimensionless control parameters and reads as
\begin{equation}
\label{Deltah0m-plates}
H_0^m \simeq [\bar{h}_J+\beta(\delta)]\, \overline{V}, \quad \overline{V} \lesssim \overline{V}_{T},
\end{equation}
where $\beta\approx \alpha(3 x/4)$ is shown in the inset of Fig.~\ref{FigS-plates}C and $\alpha$ given by Eq.~(\ref{alpha-eq}).

Fig.~\ref{FigS-plates}D shows a good collapse of the experimental data onto the theoretical prediction when the data for $h_0^m-h_J$ are rescaled by $L_0$ and plotted as a function of the new dimensionless group identified in Eq.~(\ref{Deltah0m-plates}). The grey area in Fig.~\ref{FigS-plates}D, where essentially all the data lie, highlights the region spanned by the theoretical variation of $H_0^m$ when the parameters $\bar{h}_J$ and $\delta$ are varied within their experimental range. The solid green curve shows the theoretical evolution of $H_0^m$ when the experimental average value of $\bar{h}_J$ and $\delta$ is used.

\subsection{Optimal plate distance for fluid capture by viscous entrainment}

From the analysis performed for brushes in the main text, one could think that there is no optimal geometry when two parallel plates are removed from a bath because the porosity is equal to 1. However, increasing the distance $2d$ between the plates increases the volume available for the liquid but also increases the permeabilities, $k_{\parallel}=k_{\perp}=d^2/3$, which reduces the height reached by the interface. Here again, the interplay between these two antagonistic contributions leads to a non-monotonic variation of the volume of liquid captured by viscous entrainment when $d$ is varied. To gain quantitative insight into this mechanism, we rewrite the two $d$-dependent parameters of Eq.~(\ref{h-eq-adim-final-plate2}) as follows
\begin{equation}
\label{new-parameters-plates}
\overline{V} = (\tilde{d}/d)^2 = 1/\bar{d}^{\, 2}, \quad \bar{h}_J = h^{\star}_J/\bar{d},
\end{equation}
with $\tilde{d}^{\, 2} = 3\mu V/(\rho g)$, $h_J = \ell_c^2 \cos \theta_Y /d$, $\bar{h}_J = h_J /L_0$ and $h^{\star}_J=\ell_c^2 \cos \theta_Y/(\tilde{d} L_0)$. The volume of liquid, $\mathcal{V}$, captured at the end of retraction, $\bar{t}=1$, is written as 
\begin{equation}
\label{Volume-plates}
\overline{\mathcal{V}} = \frac{\mathcal{V}}{\mathcal{V}_I} = 1+ \frac{\bar{d}}{h^{\star}_J} \int_0^1 H(\bar{y},1)  d\bar{y},
\end{equation}
where $\mathcal{V} = 2d \int_{-W/2}^{W/2} h(y,L_0/V) \, dy$, $\bar{y}=2y/W$ and $\mathcal{V}_I = 2dW h_J = 2W \ell_c^2 \cos \theta_Y$ is the volume initially between the two plates above the liquid bath level. Consequently, $H$ and thus $\overline{\mathcal{V}}$ depend only on $\bar{d}$ when $\delta$ and $h^{\star}_J$ are fixed.

Fig.~\ref{FigS-plates}E shows the non-monotonic variation of $\overline{\mathcal{V}}$ as a function of $\bar{d}$ for given values of $\delta$ and $h^{\star}_J$. Fig.~\ref{FigS-plates}F shows the variation of $\bar{d}_{\text{opt}}$, corresponding to the maximum of $\overline{\mathcal{V}}$ in Fig.~\ref{FigS-plates}E, as a function of $h^{\star}_J$ for several value of $\delta$. $\bar{d}_{\text{opt}}$ does not depend on $h^{\star}_J$ when it is small and varies weakly with $\delta$, see Inset Fig.~\ref{FigS-plates}F. For large $h^{\star}_J$, $\bar{d}_{\text{opt}}$ does not depend on $\delta$ and varies as $(h^{\star}_J)^{1/3}$, or in dimensional units ($\theta_Y=0$)
\begin{equation}
2d_{\text{opt}}  \underset{h^{\star}_J \ll 1}{=} \frac{\sqrt{3}\, \tilde{d}}{ (1+0.075\, \delta^{3/2})^{2/3}} = \frac{3\, (\mu V/\rho g)^{1/2}}{ (1+0.075\, \delta^{3/2})^{2/3}}, \quad
2d_{\text{opt}} \underset{h^{\star}_J \gg 1}{=} 1.88\, \tilde{d}\, (h^{\star}_J)^{1/3} = 2.71\, \left(\frac{\mu V \ell_c^2}{\rho g L_0}\right)^{1/3}.
\end{equation}

\end{document}